\documentclass[intlimits,twoside,a4paper]{article}

\usepackage[cp1251]{inputenc}

\usepackage[eqsecnum]{cmpj3}

\usepackage{bm}
\usepackage[makeroom]{cancel}
\usepackage{color}
\usepackage{xcolor}
\usepackage{amstext}
\usepackage[english]{babel}
\usepackage{hyperref}
\usepackage{IEEEtrantools}
\usepackage{adjustbox}

\newcommand{\h}[1]{\hat{#1}}

\newcommand{\HV}[1]{\hat{\vec{#1}}}
\newcommand{\beq}{\begin{equation}}
\newcommand{\eeq}{\end{equation}}
\newcommand{\ga}{\alpha}

\newcommand{\gz}{\zeta}
\newcommand{\gga}{\Gamma}
\newcommand{\chf}{confluent hypergeometric functions}
\newcommand{\che}{confluent hypergeometric equation}
\newcommand{\ud}{\, \mathrm{d}}
\newcommand{\mbi}{\mathbb{Z}}

\newcommand{\wt}{\widetilde{M}(a,b,z)}
\newcommand{\wta}{\widetilde{M}}

\newcommand{\blsq}{\blacksquare}
\newcommand{\bltr}{\blacktriangle}

\issue{2022}{25}{3}{33203}
\doinumber{10.5488/CMP.25.33203}

\title[A physicist's guide to the solution of Kummers equation and confluent $\cdots$]{A physicist's guide to the solution of Kummer's equation and \chf}

\author[W. N. Mathews Jr., M. A. Esrick, Z. Y. Teoh, J. K. Freericks]{W. N. Mathews  Jr.\orcid{0000-0003-3111-4265}\refaddr{label1}\thanks{W. N Mathews  Jr. passed away after completing this work, but before publication.}, {M. A.} {Esrick}\orcid{0000-0001-7686-4276}\refaddr{label1}, {Z. Y.} {Teoh}\orcid{0000-0003-2857-2417}\refaddr{label2}, {J. K.} {Freericks}\orcid{0000-0002-6232-9165}\refaddr{label1} }

\addresses{
	\addr{label1} Department of Physics, Georgetown University, 37th and O Sts. NW, Washington, DC, USA 20057-0995
	\addr{label2} Department of Mathematics, Xiamen University Malaysia, Jalan Sunsuria, Bandar Sunsuria, Sepang, 43900, Selangor, Malaysia
}

\date{Received July 04, 2022}

\begin{document}
\maketitle


\begin{abstract}
The confluent hypergeometric equation, also known as Kummer's equation, is one of the most important differential equations in physics, chemistry, and engineering. Its two power series solutions are the Kummer function, $M(a,b,z)$, often referred to as the confluent hypergeometric function of the first kind, and $\wt\equiv z^{1-b}\;M(1+a-b,2-b,z)$, where $a$ and $b$ are parameters that appear in the differential equation.  A third function, the Tricomi function, $U(a,b,z)$, sometimes referred to as the confluent hypergeometric function of the second kind, is also a solution of the \che\;that is routinely used.  Contrary to common procedure, all three of these functions (and more) must be considered in a search for the two linearly independent solutions of the \che.\;  There are situations, when $a, b$, and $a - b$ are integers, where one of these functions is not defined, or two of the functions are not linearly independent, or one of the linearly independent solutions of the differential equation is different from these three functions. Many of these special cases correspond precisely to cases needed to solve problems in physics.  This leads to significant confusion about how to work with confluent hypergeometric equations, in spite of authoritative references such as the NIST Digital Library of Mathematical Functions. Here, we carefully describe all of the different cases one has to consider and what the explicit formulas are for the two linearly independent solutions of the confluent hypergeometric equation. The procedure to properly solve the confluent hypergeometric equation is summarized in a convenient table.  As an example, we use these solutions to study the bound states of the hydrogenic atom, correcting the standard treatment in textbooks. We also briefly consider the cutoff Coulomb potential.  We hope that this guide will aid physicists to properly solve problems that involve the confluent hypergeometric differential equation.


\end{abstract}


\maketitle

\section{Introduction\label{S:Introduction}}

We are taught how to solve second-order linear differential equations early in our study of physics. The procedure is straightforward, but sometimes may be complicated to carry out. We identify the two linearly independent solutions, and then use either initial conditions, or boundary conditions to select the proper solution being sought. However, when one works with equations of the hypergeometric type (and here, we focus on the confluent hypergeometric equation), there is no general way to identify the two linearly independent solutions for all values of the parameters in the differential equation. This means that the general solution strategy will not work so easily, and requires extra care to be carried out correctly. This point is a subtle one, and is missed, for example, in essentially all quantum mechanics textbooks in the description of how to solve the energy eigenvalues and wavefunctions for the Coulomb problem of the hydrogen atom (and many other problems as well). In this work, we carefully describe how the general procedure is modified to enable solving the confluent hypergeometric equation for boundary value problems and we present the proper (and complete) treatment of the Coulomb problem for hydrogen~\cite{DLMF,AS,Slater}.

We begin with the basic definitions of the Kummer and Tricomi functions, $M(a,b,z)$ and $U(a,b,z)$, respectively.  We note that, contrary to more-or-less common practice, \textit{the two power series solutions of Kummer's equation (also known as the \che), namely $M(a,b,z)$ and $\wt$, defined in equation~\eqref{Eq:wtM}, absolutely must be included in the considerations}.  We discuss in detail the circumstances in which these three solutions are, and are not, defined, and are, and are not, linearly independent, emphasizing the complicated ways in which the characters of $a$ and $b$ and the constraints on $a-b$ are all-important.  We stress the great care that is needed in determining how these three solutions are to be used to obtain two linearly independent solutions of Kummer's equation and the circumstances in which they cannot, where additional functions must be used.

The \che, or Kummer's equation, is given by
\beq \label{Eq:che1}
z\frac{\ud^2 w}{\ud z^2} + (b - z) \frac{\ud w}{\ud z} - a w = 0,
\eeq 
with $a$ and $b$ constants. This differential equation is in the Laplace form~\cite{Schlesinger,Ince}, where the coefficients of the different terms are at most linear functions in $z$, although we will not be using the Laplace method in this work. The confluent hypergeometric equation is an important differential equation that is used in many areas of classical and quantum physics, chemistry, and engineering~\cite{Seaborn}.  The underlying reason for this importance is that many of the special functions of mathematical physics can be expressed in terms of \chf\;and many of the differential equations of physics, chemistry, and engineering can be reduced to the \che\;and thus solved in terms of \chf.  This is particularly true for quantum mechanics~\cite{Seaborn,MF,Merzbacher,LandL,Flugge,Negro et al.,Williams,Dong,Pena et al.,AWH,Puri,Othman et al.}, where, for example, the bound state problems for the simple harmonic oscillator in one, two and three dimensions, the Coulomb problem in two and three dimensions, and the Cartesian one-dimensional Morse potential can all be solved in terms of confluent hypergeometric functions.  In addition, continuum problems, such as the free particle in one, two, and three dimensions, the one-dimensional Cartesian linear potential, the continuum of the Coulomb problem in two and three dimensions, and the continuum of the Cartesian one-dimensional Morse potential, can also be solved using confluent hypergeometric functions.  Moreover, we note that there is a very nice discussion of Landau levels that also employs \chf\;\cite{Ciftja}.  The \che\; also arises in optics~\cite{AWH,Galue,Tang,Jin,Augustyniak}, classical electrodynamics~\cite{Seaborn,Galue,Georgiev}, classical waves~\cite{MF,Whitham,Liu et al.}, diffusion~\cite{Kalla}, fluid flow~\cite{Koppel}, heat transfer~\cite{Jeong}, general relativity~\cite{Black Hole,Li,Turyshev et al.,Bero}, semiclassical quantum mechanics~\cite{Silverstone}, quantum chemistry~\cite{Rahman,Montgomery et al.}, graphic design~\cite{Inoguchi et al.}, and many other areas.  The solutions of the \che\;depend in an essential way on whether or not $a, b$, and $a - b$ are integers and the standard references (see below) do not present these solutions, with appropriate qualifications, in a user-friendly way.  

The primary purpose of this paper is to properly organize the solutions of the \che, so as to allow one to navigate the challenging and convoluted labyrinth of possible combinations of $a$ and $b$, and to discuss the associated subtleties.  Our principal results in this regard are summarized in table~\ref{Table:Labyrinth} in section \ref{S:Section3}.  We expect this table to be very useful in determining the correct solutions of the \che~for problems in physics, other sciences, engineering, and mathematics.  We also expect that working through the analysis in the Appendix that results in table~\ref{Table:Labyrinth} would go a long way toward relieving any unfamiliarity with \chf.

A comprehensive discussion of the history of the hypergeometric function, from which confluent hypergeometric functions are descended, has been written by Dutka~\cite{Dutka}.  In particular, unlike most other special functions, which were defined as the solutions of their corresponding differential equation, the hypergeometric functions were first defined in terms of their power series, and the differential equation that they satisfy was discovered later. Since the hypergeometric functions are not defined or are not distinct for some integer values of their parameters, this introduces challenges with describing all of the linearly independent solutions of the corresponding differential equation. This difficulty spills over to the confluent hypergeometric functions, which are a special case of the hypergeometric functions. It turns out that for many physics applications, we need the solutions of the confluent hypergeometric equation precisely for cases where $a, b$, or $a-b$ are integers where the analysis becomes more nuanced.  

Our principal references on the \chf\;are the NIST Digital Library of Mathematical Functions (DLMF)~\cite{DLMF}, the precursor volume, \textit{The Handbook of Mathematical Functions}, by Abramowitz and Stegun (AS)~\cite{AS}, \textit{Confluent Hypergeometric Functions}, by Slater~\cite{Slater}, and \textit{Higher Transcendental Functions}, edited by Erd\'{e}lyi~\cite{Bateman}.  Some other useful sources of information about \chf\;are \textit{Mathematical Methods for Physicists}, by Arfken,  Weber, and Harris~\cite{AWH},  \textit{Methods of Theoretical Physics}, by Morse and Feshbach~\cite{MF}, \textit{ A Course of Modern Analysis}, by Whittaker and Watson~\cite{WW}, \textit{Special Functions in Physics with MATLAB}, by Schweizer~\cite{Schweizer}, the Wolfram MathWorld website~\cite{Weisstein 1, Weisstein 2,Weisstein 3,Weisstein 4}, and a beautiful dynamic calculator of the Kummer function, $M(a,b,z)$, and the Tricomi function, $U(a,b,z)$, on the Wolfram website~\cite{Demo}.  There is also a Wikipedia entry titled ``Confluent hypergeometric functions''~\cite{Wikipedia}.  In addition, we particularly note two papers which consider the general solution of the stationary state Schr\"{o}dinger equation in terms of \chf\;\cite{Negro et al.,Pena et al.}, two textbooks which employ \chf\;in a discussion of the bound and continuum states of the hydrogen atom  and other problems in quantum mechanics~\cite{LandL,Puri}, and a paper which considers the use of \chf\; in determining the bound states of the attractive Coulomb potential~\cite{Othman et al.}.

This paper is organized as follows.  In section \ref{S:Basics}, we present and discuss the basic definitions and properties of the three standard solutions of the \che.  In section \ref{S:Section3}, we present table I in which the linearly independent solutions of the \che\;are organized according to the possible values of $a$ and $b$ and the constraints on $a-b$, thereby imbuing the labyrinth of values of $a$ and $b$ with some order.  In section \ref{S:Section4}, we present the limiting values, as $z \to 0$ and $z \to \infty$, of the Kummer function, $M(a,b,z)$ and the Tricomi function, $U(a,b,z)$.  As a noteworthy example, for which integral values of $b$ are germane, we consider in section \ref{S:Section5} the quantum-mechanical treatment of the bound states of the hydrogenic atom.  We show that a careful and complete treatment is more complex than the standard approach found in quantum mechanics textbooks.  In section \ref{S:Section6},  as an example that emphasizes the care that must be used in working with \chf, we briefly consider the cutoff Coulomb potential discussed by Othman,  de Montigny, and Marsiglio~\cite{Othman et al.}, illustrating some of the subtle issues not discussed in their work.  In section \ref{S:Section7}, we provide our conclusions.  In the Appendix, we present the detailed analysis that results in table~\ref{Table:Labyrinth}.

\section{Basic definitions and properties}\label{S:Basics}

In some problems in quantum mechanics, the first index of the \che, $a$, is a non-positive integer .  For example, in solving the Schr\"{o}dinger equation for the bound states of hydrogen, we find that $a = \ell+1-n$, where $n>\ell$ is the principal quantum number and the eigenvalues of $\HV{L}^{2}$, where $\HV{L}$ is the orbital angular momentum operator, are $\ell(\ell + 1)\hbar^2$, and $\ell$ is a non-negative integer. In this instance, that is, when $a$ is a non-positive integer, $U(a,b,z)$ is a polynomial in $z$, and provided $b$ is not a non-positive integer, $M(a,b,z)$ exists and $U(a,b,z) \propto M(a,b,z)$.
 
In addition, it frequently occurs in applications of \chf\; that the second index of $M(a,b,z)$ and $U(a,b,z)$, $b$, is an integer.  For example: when we solve the Schr\"{o}dinger equation in plane polar coordinates, $b$ can take on the values $1-2|\tilde{m}|$ and $1+2|\tilde{m}|$, where the eigenvalues of $\h{L}_z$, the $z$-component of the orbital angular momentum operator, are $\tilde{m}\hbar$, and $\tilde{m}$ is an integer; and when we solve the Schr\"{o}dinger equation in spherical coordinates, $b$ can take on the values $-2\ell$ and $2(\ell+1)$ (we use $\tilde{m}$ because we are reserving the symbol $m$ for another purpose). 

In what follows, we will use the symbols $\mathbb{Z},~ \mathbb{Z}^{\leqslant 0},~\mathbb{Z}^{> 0}$, and $\mathbb{Z}^{\geqslant 2}$ to designate the sets of integers, non-positive integers, positive integers, and integers $\geqslant 2$, respectively.  Furthermore, the symbol $\in$ means ``is in'' or ``belongs to'', the symbol $\not \in$ means ``is not in'' or ``does not belong to'', and the symbol $\forall$ means ``for all''.

When $b \in \mathbb{Z}$, there are three classes of problems with regard to the solutions of the \che.  To reveal these problems, we consider the standard Frobenius (generalized power series) method of solution for a linear, ordinary differential equation, which is valid for an expansion about a point which is a regular point or a regular singular point~\cite[\S 7.5]{AWH} and~\cite{Riley et al.,McQuarrie,Boas,Frobenius} of the differential equation.  Since the \che\;has a regular singularity at $z = 0$, we can attempt a solution of the form given by 
\beq \label{Eq:w1}
w(a,b,z) = \sum_{s=0}^{\infty} C_s z^{\lambda + s},
\eeq
where $\lambda$ is a pure number to be determined.  Substitution of this putative solution into the \che\;reveals that the two possible values of $\lambda$ are $0$ and $1-b$.   The corresponding first and second power series solutions of the \che\;are denoted by $M(a,b,z)$ and $\wt$ [which is defined in equation \eqref{Eq:wtM}], respectively.  Here, $M(a,b,z)$ is Kummer's function, which is sometimes referred to as the confluent hypergeometric function of the first kind, and is also denoted by $_1F_1 (a;b;z)$ and $_1F_1[a;b;z]$ [see 13.1.10 of AS  and equation (1.1.7) of Slater~\cite{Slater2}, respectively].  Its power series definition is given by
\beq \label{Eq:M}
M(a,b,z) = \sum_{s=0}^\infty \frac{(a)_s}{(b)_s}\; \frac{z^s}{s!}.
\eeq
The Pochhammer symbol (which is also known as the Pochhammer function, the Pochhammer polynomial, the rising factorial, the rising sequential product, and the upper factorial)~\cite{Wikipedia,Pochhammer} is given by
\beq \label{Eq:PochhammerSymbol}
(a)_0 = 1, (a)_1 = a,\quad \text{and}\; (a)_s = a(a+1) \ldots (a+s-1) = \frac{\Gamma (a+s)}{\Gamma (a)}, \quad \text{for}\;s \in \mathbb{Z}^{>0},
\eeq
where $\Gamma$ denotes the standard gamma function (see 13.2.2 of DLMF and 13.1.2 of AS). The Kummer function, $M(a,b,z)$, is an entire function of $z$ and $a$, and is a meromorphic function of $b$  (see 13.2.4 of DLMF).

Some additional information about the Pochhammer symbol may be useful. In this list, $m,\, n,\, s,\, m-s \in \mathbb{Z}^{\geqslant0}$.
\begin{enumerate}
    \item $(a)_s$ is defined if and only if $s \in \mathbb{Z}^{\geqslant0}$.  Morever,\;$a$ can be any real or complex number.
    \item For $a \not\in \mathbb{Z}^{\leqslant0}$, $(a)_s \ne 0$. 
    \item For $s \geqslant m+1$, $(-m)_s = 0$. 
    \item For $n < s \leqslant m,~ (-n+s)_{m-s} = \frac{(-n+m-1)!}{(-n+s-1)!}$, which is       also valid in the limit $n \to s$.
    \item For $m > n \geqslant s,~ (-n+s)_{m-s} = 0$,
    \item For $n \geqslant m \geqslant s$, $(-n+s)_{m-s} =         (-1)^{m+s} \frac{(n-s)!}{(n-m)!}$. 
    \item For $m \geqslant s$ and $a \not \in \mbi$, or $a > m \geqslant s$ and $a \in \mbi^{\geqslant 0}$, $ (1-a+s)_{m-s} = \frac{(1-a)_m}{(1-a)_s}$.
    \item Since $(a)_s = \frac{\gga(a+s)}{\gga(a)}, (a)_s = 0$ if $a \in \mathbb{Z}^{\leqslant0}$ and          $a+s \in \mathbb{Z}^{>0}$.  Also, $a \in \mathbb{Z}^{\leqslant0}$ and $a+s \in 
          \mathbb{Z}^{\leqslant0} \Longrightarrow (a)_s$ is indeterminate.  
\end{enumerate}    

In addition, in table 2.1 on page 19, Seaborn~\cite{Seaborn} gives several identities involving Pochhammer symbols.

From here on, because $z^{1-b}\;M(1+a-b,2-b,z)$ occurs so frequently, and because $M(a,b,z)$, $z^{1-b}\;M(1+a-b,2-b,z)$, and $U(a,b,z)$ should be regarded as an essentially the same footing as far as solutions of Kummer's equation,  we define 
\beq \label{Eq:wtM}
\wt \equiv z^{1-b}\;M(1+a-b,2-b,z);
\eeq
that is, we use $M(a,b,z)$ and $\wt$ for, respectively, the first and second power series solutions of the \che.

Using $M(a,b,z)$ as given by equation \eqref{Eq:M}, we can immediately see the three classes of problems that occur when $b \in \mathbb{Z}$.  First, from equations \eqref{Eq:M} and \eqref{Eq:PochhammerSymbol}, when $b \in \mathbb{Z}^{\leqslant 0}$, $(b)_s$ is $0$ for some value of $s \in \mathbb{Z}^{\geqslant 0}$. This means that $M(a,b,z )$ is not defined for $b \in \mathbb{Z}^{\leqslant 0}$.  (Alternatively, it can be said that $M(a,b,z)$, regarded as a function of $b$, has simple poles for $b \in \mathbb{Z}^{\leqslant 0}$~\cite{Slater}.)  Second, if $b = 1$, then $M(a,b,z) = \wt$, i.e., the two power series solutions of Kummer's equation are the same.  Third if $b \in \mathbb{Z}^{\geqslant 2}$, then $\wt$ is not defined.  Since the \che\;is a second order, linear, ordinary, differential equation, and it has no new singular behavior when $b \in \mathbb{Z}$, it must have two linearly independent solutions even when $b \in \mathbb{Z}$. The key point that we emphasize here is that \textit{the fact that one of the \chf\;is not defined does not mean that the differential equation no longer has two linearly independent solutions}. What it means is that the two linearly independent solutions must be determined with care. This is a point that can be easily misunderstood and which can lead to erroneous conclusions when solving problems that reduce to the confluent hypergeometric equation.

According to 13.2.3 of DLMF, ``$M(a,b,z)$ does not exist when $b$ is a non-positive integer''.  However, AS  includes in 13.1.3 a short table about the character of $M(a,b,z)$, which includes an explicit indication that $M(a,b,z)$ can be defined when $b \in \mathbb{Z}^{\leqslant 0}$, and the circumstances under which it is not defined. We consider the entries in this table, other than the first two and the last, to be dubious.  Moreover, there are other entries in Chapter 13 of AS, particularly 13.6.2 and 13.6.5, that can be interpreted as indicating that $M(a,b,z)$ with $b \in \mathbb{Z}^{\leqslant 0}$ can be defined.  Reference~\cite{MF} also indicates, albeit somewhat indirectly, on page 605, that $M(a,b,z)$ is not defined if $b \in \mathbb{Z}^{\leqslant 0}$.  Reference~\cite{AWH} explicitly states on page 917 that $M(a,b,z)$ is not defined if $b \in \mathbb{Z}^{\leqslant 0}$.  Reference~\cite{Slater}  also indicates on pages 2 and 3 that $M(a,b,z)$ is not defined for $b \in \mathbb{Z}^{\leqslant 0}$.  In \S\;6.7.1, of~\cite{Bateman}, it is noted that $M(a,c,z)$ `` \ldots fails to be defined at $c=0, -1, -2, \ldots $''.  Last, but not least, what is written on pages 347 and 348 of~\cite{WW} can be interpreted as stating that $M(a,b,z)$ is not defined if $b \in \mathbb{Z}^{\leqslant 0}$.

The full hypergeometric function, or just the hypergeometric function, $F(a,b;c;z)$ (see Chapter 15 of DLMF and Chapter 15 of AS), also is not defined when $c \in \mathbb{Z}^{\leqslant 0}$. Both DLMF and AS  discuss alternate solutions in this situation.

In 13.2.2 and 13.2.3, DLMF presents the Kummer function, $M(a,b,z)$, and Olver's function, as ``The first two standard solutions'' of Kummer's equation.  It would appear that the solution to the problem that Kummer's function, $M(a,b,z)$, is not defined for $b \in \mathbb{Z}^{\leqslant 0}$, is simply to instead use Olver's function.  Unfortunately, Olver's function is not always a non-trivial solution of the \che.  Accordingly, we shall not use Olver's function in our considerations and thus we say no more about it.

The second standard solution of the \che\;is often taken to be the Tricomi function, today generally denoted by $U(a,b,z)$, and sometimes denoted as the solution of the second kind.  This solution can be defined, when $b \not\in \mathbb{Z}$, as a linear combination of the two power series solutions of the \che,  according to
\beq \label{Eq:U1}
U(a,b,z) = \frac{\gga(1-b)}{\gga(1+a-b)}\;M(a,b,z) + \frac{\gga(b-1)}{\gga(a)}\;z^{1-b}\;M(1+a-b,2-b,z), \quad b \not\in \mathbb{Z}.
\eeq
(See 13.2.42 of DLMF and \S 1.3 of~\cite{Slater}; this relation is not given in AS.)  More generally, $U(a,b,z)$ is the solution defined uniquely by the property
\beq \label{Eq:U2}
U(a,b,z) \sim z^{-a}, \quad \text{as}\;z \to \infty,\; \quad \text{for}\; -\piup < \arg{z} <  \piup\;.
\eeq
(See 13.2.6 of DLMF and 13.1.8 of AS.)  The function $U(a,b,z)$ has a branch point at $z = 0$, and we choose the principal branch to have a branch cut along $(-\infty,0]$, corresponding to the principal branch of $z^{-a}$.  [See 13.2.6 of DLMF.  Note that $\arg(z)$ refers to the \textit{phase} of the generally complex number, $z$, and that DLMF  uses ph instead of $\arg$.]

Since (see 5.5.3 of DLMF and 6.1.17 of AS)
\beq
\gga(u)\;\gga(1-u) = \frac{\piup}{\sin(\piup u)}, \quad \text{for}\;u \not\in \mathbb{Z},
\eeq
we can  use equation~\eqref{Eq:U1} to write for $b \not\in \mathbb{Z}$
\beq \label{Eq:U3}
U(a,b,z) = \frac{\piup}{\sin(\piup b)}\Bigg[\frac{M(a,b,z)}{\gga(1+a-b)\;\gga(b)} - z^{1-b}\;\frac{M(1+a-b,2-b,z)}{\gga(a)\;\gga(2-b)} \Bigg].
\eeq
This relation is given as 13.1.3 in AS, equation (1.3.5) in Slater\cite{Slater}, and equation 18.4.2 of Arfken et al.~\cite{AWH}, but is not given in DLMF.  More to the point, AS and Slater~\cite{Slater} assert that it can be defined in the limit as $b$ approaches an integer, although neither shows the corresponding calculation.   

To be clear, we indicate explicitly how $U(a,b,z)$ can be obtained:  
\begin{enumerate}  
    \item For $b \not\in  \mathbb{Z},\;U(a,b,z)$ is given by equation \eqref{Eq:U1} or \eqref{Eq:U3}, or DLMF 13.2.42, or AS 13.1.3.  If $a \in \mbi^{\leqslant 0}$, we can also use DLMF 13.2.7.  
    \item For $b \in  \mathbb{Z}^{\leqslant0}$:
    \begin{itemize}
        \item For $a \not \in \mbi$, we use DLMF 13.2.11 followed by DLMF 13.2.9, or we can also        use DLMF 13.2.30.
        \item For $a \in \mathbb{Z}^{\leqslant0}$, we can use DLMF 13.2.7.
        \item For $a \in \mbi^{\leqslant 0}$ and $a \geqslant b$, or equivalently, $a \in \mbi^{\leqslant 0}$ and         $a-b \ne - (1+q)$ where $q \in \mbi^{\geqslant 0}$, we can also use DLMF 13.2.7 or DLMF         13.2.32.
        \item For $a \in \mbi^{\leqslant 0}$ and $a < b$, or equivalently, $a \in \mbi^{\leqslant 0}$ and $a-b       = -(1+n)$ where $n \in \mbi^{\geqslant 0}$, we can use DLMF 13.2.7 or DLMF 13.2.8.  (In         discussing the constraints on $a - b$ elsewhere, we use $q$ instead of $n$; we use       $n$ here only because DLMF 13.2.8 does.)
        \item For $a \in \mbi^{> 0}$, we can use DLMF 13.2.11 followed by DLMF 13.2.9, or we can        use DLMF 13.2.30.
        \item Anytime DLMF 13.2.7 is used and $b \in \mbi^{\leqslant 0}$, the contents between the two        $=$'s must be deleted, since $M(a,b,z)$ is not defined for $b \in \mbi^{\leqslant 0}$.
    \end{itemize}
    \item For $b \in \mathbb{Z}^{> 0}$:
    \begin{itemize}
        \item For $a \not \in \mbi$, we can use DLMF 13.2.9 or DLMF 13.2.27.
        \item For $a \in \mathbb{Z}^{\leqslant 0}$, we can use DLMF 13.2.7 or DLMF 13.2.10F.
        \item For $a \in \mbi^{> 0}$ and $a \geqslant b$, or equivalently $a \in \mbi^{> 0}$ and $a-b         \ne -(1+q)$ where $q\in \mbi^{\geqslant 0}$, we can use DLMF 13.2.9 or DLMF 13.2.27.
        \item For $a \in \mbi^{> 0}$ and $a < b$, or equivalently $a \in \mbi^{> 0}$ and $a-b           = -(1+q)$ where $q\in \mbi^{\geqslant 0}$, we can use DLMF 13.2.9 or DLMF 13.2.29.
    \end{itemize}       
     \item For $a \in \mathbb{Z}^{\leqslant0}~$, and $\forall b, U(a,b,z)$ is given by DLMF 13.2.7.           Of course, for $b \in \mathbb{Z}^{\leqslant0}$, the contents between the two $=$'s in DLMF         13.2.7 must be deleted.
    \item 13.2.27 of DLMF (with $b = 1+n$ and $a-n \ne -q$, where $q \in \mathbb{Z}^{\geqslant 0})             \Longrightarrow (-1)^n n!\;\gga(a-n) U(a,1+n,z)$.
    \item 13.2.29 of DLMF (with $a=1+m, b=1+n, m \in \mathbb{Z}^{\geqslant 0}, n \in \mathbb{Z}^{\geqslant 0},         m<n) \Longrightarrow$ \\ $\frac{m!}{(n-m-1)!}\;U(1+m,1+n,z)$. 
    \item 13.2.30 of DLMF (with $a \not \in \mbi^{\leqslant 0}b=-n$, $n \in \mathbb{Z}^{\geqslant 0}$)                 $\Longrightarrow (-1)^{n+1}(n+1)!~\gga(a)~U(a,-n,z)$.
    \item 13.2.32 of DLMF (with $a=-m,b=-n, m \in \mathbb{Z}^{\geqslant 0}, n \in \mathbb{Z}^{\geqslant 0},         m\leqslant n) \Longrightarrow \frac{(n-m)!}{m!}\;U(-m,-n,z)$.
    \item DLMF 13.2.28 and 13.2.31 do not yield Tricomi functions.
    \item As long as $a-n \not\in \mathbb{Z}^{\leqslant0}$, 13.2.9 of DLMF  contains $\ln z$ terms. 
          DLMF 13.2.27, 13.2.28, 13.2.30, and 13.2.31 also contain $\ln z$  terms.
    \item AS 13.1.6 does yield DLMF 13.2.9, but AS 13.1.6, apparently inadvertently, omits the requirement that $a \not \in        \mbi^{\leqslant 0}$.
\end{enumerate}
Much of this is noted in table~\ref{Table:Labyrinth}.

Just as there are issues with the Kummer function, $M(a,b,z)$, so there are also issues with the Tricomi function, $U(a,b,z)$.  When $a \in \mathbb{Z}^{\leqslant 0}$ and $b \not\in \mathbb{Z}^{\leqslant 0}$, $U(a,b,z)$ is proportional to $M(a,b,z)$, i.e., the Tricomi function is proportional to the first power series solution. (See 13.2.7 and 13.2.10 of DLMF.)  In addition, when $a - b = - (1 + q)$, where $q \in \mathbb{Z}^{\geqslant 0}$ (or equivalently $1+a-b \in \mathbb{Z}^{\leqslant 0}$), and $2-b \not\in \mathbb{Z}^{\leqslant0}$, then $U(a,b,z)$ is proportional to $\wt$, i.e., the Tricomi function is proportional to the \textit{second} power serises solution (see 13.2.8 of DLMF).  An additional complication is the occurrence of solutions of the \che\;which contain logarithmic terms.  In \S 6.7.1,~\cite{Bateman} notes that ``Whenever $c$ is an integer'', $M(a,c,z)$ and $M(a-c+1,2-c,z)$ ``provide one solution, and the second solution will contain logarithmic terms''.  It appears that this is not always true.  As we will see, in only six of the eight cases where $b \in \mathbb{Z}$ is there a $\ln z$ term; in Case 1.B, Case 1.C, Case 5.B, and Case 5.C, the $\ln z$ terms enter via $U$; in Case 4.B and Case 4.C, the $\ln z$ terms enter via the non-standard second solution; in Case 3.B and 6.C, $U$ is one of the solutions and yet there are no $\ln z$ terms.  Reference~\cite{Slater} also carefully discusses, in \S 1.5 and \S 1.5.1, the ``logarithmic solutions when $b$ is an integer''.  Of course, such solutions are usually not compatible with the boundary conditions appropriate for most problems in physics.  

We are going to be more or less continually concerned with the circumstances under which $M(a,b,z), \\  \widetilde{M}(a,b,c)$, and $U(a,b,z)$ exist and whether we have two linearly independent solutions.  The salient facts are as follows:
\begin{enumerate}
	\item When $b \in \mbi, M(a,b,\gz)$ and $\wt \Longrightarrow$ only one      
	      solution.  Specifically:
	\begin{itemize}  
		\item $b \in \mbi^{\leqslant 0} \Longrightarrow M(a,b,\gz)$ is not defined [see equation \eqref{Eq:M}].
		\item $b = 1 \Longrightarrow \wt = M(a,b,z)$ [see equation \eqref{Eq:wtM}].
	    \item $b \in \mbi^{\geqslant 2}  \Longrightarrow \wt$ is not defined [see equations \eqref{Eq:M} and \eqref{Eq:wtM}].
	\end{itemize}
\item $a \in \mathbb{Z}^{\leqslant 0}$, $b \not\in \mathbb{Z}^{\leqslant 0} \Longrightarrow U(a,b,z) \propto M(a,b,z)$ [$b \not\in \mathbb{Z}^{\leqslant0} \Longrightarrow M(a,b,z)$ is defined. See 13.2.7 of DLMF.].
\item $a - b = - (1+q)$ or equivalently $b = 1+a+q$ with $q \in \mathbb{Z}^{\geqslant 0}$, and $1-a-q \not\in \mathbb{Z}^{\leqslant 0} \Longrightarrow U(a,b,z) \propto \wt\mid_{b\,=\,1+a+q} = z^{-(a+q)}\;M(-q,1-a-q,z)$  [$1-a-q \not\in \mathbb{Z}^{\leqslant 0} \Longrightarrow M(-q,1-a-q,z)$ is defined.  See 13.2.8 of DLMF.]. 
\item $a \not\in \mathbb{Z}^{\leqslant 0}, b \not\in \mathbb{Z}^{\leqslant 0} \Longrightarrow  M(a,b,z)$ and $U(a,b,z)$ are linearly independent solutions  [$a \not\in \mathbb{Z}^{\leqslant 0} \Longrightarrow U(a,b,z)\;\not\propto M(a,b,z), b \not\in \mathbb{Z}^{\leqslant 0} \Longrightarrow M(a,b,z)$ is defined]. 
\item $b \not\in \mathbb{Z} \Longrightarrow  M(a,b,z)$ and $\wt$ are linearly independent solutions  [$b \not\in\mathbb{Z}^{\leqslant 0} \Longrightarrow M(a,b,z)$ is defined, $b \ne 1 \Longrightarrow M(a,b,z) \ne \wt, b \not \in \mathbb{Z}^{\geqslant 2} \Longrightarrow \wt$ is defined].
\item $b \not \in \mathbb{Z}^{\geqslant 2}, b \ne 1+a+q$ with $q \in \mathbb{Z}^{\geqslant0} \Longrightarrow U(a,b,z)$ and $\wt$ are linearly independent solutions.  [$b$ not an integer $\geqslant 2 \Longrightarrow \wt$ is defined, $b \ne 1+a+q \Longrightarrow U(a,b,z) \not\propto \wt$].
\item $U(a,b,z) = z^{1-b}\;U(1+a-b,2-b,z)$.  This is the second of the Kummer transformations (see 13.2.40 of DLMF and 13.1.29 of AS).
\end{enumerate} 
The first three points are particularly crucial and must be kept in mind at all times; they tell us when at most two of the three standard solutions are available for linearly independent solutions. The primary implication of point 7 is that while determining whether $M(a,b,z)$ and $\wt$ are linearly independent solutions is straightforward, determining two forms of $U$ that are linearly independent solutions is not straightforward.  Such information is given by 13.2.24 and 13.2.25 of DLMF, and 13.1.16--13.1.19 of AS, but this information is apparently not needed for our purposes.

The constraints $a-b \ne - (1+q)$ or $b \ne 1+a+q$, and $a-b = - (1+q)$ or $b = 1+a+q$, where $q \in \mathbb{Z}^{\geqslant 0}$, complicate matters.  Let us consider the second requirement, $b = 1+q+a$ or $a-b = - (1+q)$.
\begin{itemize}
\item From points 1 and 2 above, we see that, with $a = - m$, $m \in \mathbb{Z}^{\geqslant 0}, b \in \mathbb{Z}^{> 0}$, both $M(-m,b,z)$ and $U(-m,b,z)$ exist, but are not linearly independent solutions.  This is relevant because $a = - m$ with $m \in \mathbb{Z}^{\geqslant 0}$ and $b \in \mathbb{Z}^{> 0}$ ensure that $a - b = - (1+q)$ is satisfied.
\item From point 3, we see that $a-b = - (1+q)$ or $b = 1+a+q$ and $1-a-q \not\in \mathbb{Z}^{\leqslant0}$, with $q \in \mathbb{Z}^{\geqslant 0},  \Longrightarrow  U(a,1+a+q,z)$ and $\wta\mid_{b\,=\,1+a+q}  = z^{-(a+q)}\;M(-q,1-a-q,z)$ are also not linearly independent solutions.
\item $b = 1+a+q$, with $q \in \mathbb{Z}^{\geqslant 0}$ means that $a \not\in\mathbb{Z} \Longleftrightarrow b \not\in \mathbb{Z}$, and $a \in \mathbb{Z} \Longleftrightarrow b \in \mathbb{Z}$.
\item If $a-b=-(1+q)$, with $q \in \mathbb{Z}^{\geqslant 0}$, then $a = - m \Longrightarrow b = - m + (1+q)$, in which case $b \in \mathbb{Z}^{\leqslant 0} \Longrightarrow m \geqslant 1+q$.  Thus, $a = - m$, with $m \geqslant 1+q \Longrightarrow b \in \mathbb{Z}^{\leqslant 0}$, or equivalently, $a \in \mathbb{Z}^{\leqslant 0}$ and $a \leqslant -(1+q) \Longrightarrow b \in \mathbb{Z}^{\leqslant 0}$.
\item  Moreover, if $a-b=-(1+q)$ with $q \in \mathbb{Z}^{\geqslant 0}$, then $a = - m$ and $b \in \mathbb{Z}^{> 0} \Longrightarrow m \leqslant q$, so that $a = - m$, with $m \leqslant q \Longrightarrow b \in \mathbb{Z}^{> 0}$, or equivalently, $a \in \mathbb{Z}^{\leqslant 0}$ and $a \geqslant - q \Longrightarrow b \in \mathbb{Z}^{> 0}$. 
\item If $a-b=-(1+q)$ with $q \in \mathbb{Z}^{\geqslant 0}$, then $a \in\mathbb{Z}^{\geqslant 0} \Longrightarrow b \in \mathbb{Z}^{> 0}$.  
\end{itemize}  

\section{Navigating the labyrinth of values of \textit{a} and \textit{b}}\label{S:Section3}

On the one hand, one might expect that navigation through the labyrinth of values of $a$ and $b$ would be facilitated by dividing up both $a$ and $b$ into three categories: not an integer, or equivalently, $\not\in \mathbb{Z}$; non-positive integer, or equivalently, $\in \mathbb{Z}^{\leqslant 0}$; positive integer, or equivalently, $\in \mathbb{Z}^{> 0}$.  On the other hand, if one does that for $b$, the categorization of the values of $a$ becomes more complicated.  For each of the three categories of $a$, there are two possibilities: $a-b \ne - (1+q)$, or equivalently,  $1+a-b \not \in \mathbb{Z}^{\leqslant 0}$; $a-b = - (1+q)$, or equivalently, $1+a-b \in \mathbb{Z}^{\leqslant 0}$.  These two different constraints on $a-b$ stem from the definition of $\wta$, equation~\eqref{Eq:wtM} and DLMF 13.2.8.  (Here, as elsewhere, $q \in \mathbb{Z}^{\geqslant 0}$.)

The rational for this scheme stems from the entries, near the end of the previous section, in our enumeration of the circumstances under which $M(a,b,z), \widetilde{M}(a,b,c)$, and $U(a,b,z)$ exist and whether they furnish two linearly independent solutions.

We designate the categories of $a$, i.e., the last six rows in table~\ref{Table:Labyrinth}, in this section, below, with the positive integers 1--6, as indicated in table~\ref{Table:Labyrinth}. We designate the categories of $b$, i.e., the rightmost three columns in table~\ref{Table:Labyrinth}, with the capital letters A, B, C.  We thus see that there are 18 distinct cases to be considered.  The detailed analysis of these cases, which results in table~\ref{Table:Labyrinth}, is in the Appendix. We expect that it is absolutely necessary to take the time and effort to follow and understand the reasoning of the analysis in the Appendix in order to make intelligent and efficient use of table~\ref{Table:Labyrinth}. To put it another way, working through the Appendix should sufficiently sensitize the reader to the intricacies of the \chf\;in order to facilitate the effective use of table~\ref{Table:Labyrinth}.

We use table~\ref{Table:Labyrinth} in section~\ref{S:Section5} in our discussion of the bound states of the hydrogenic atom, and in section~\ref{S:Section6} in our brief discussion of the cutoff Coulomb potenial, and we discuss it briefly in section~\ref{S:Section7}.

We emphasize that of the eight cases where $b \in \mathbb{Z}$, only six, indicated by $\blsq$ in table~\ref{Table:Labyrinth}, have solutions with $\ln z$ terms.  In Case 3.B, 13.2.7 and 13.2.32 of DLMF indicate no possibility of a $\ln z$ term in $U(a,b,z)$. In Case 6.C, 13.2.9 indicates the possibility of a $\ln z$ term in $U(a,b,z)$, but $a-n \leqslant 0$, and so there is no $\ln z$ term. 

\renewcommand{\arraystretch}{1.2}

\begin{table}[!t]
	\caption{Labyrinth of values of $a$ and $b$ for the solutions of the \che.  The symbols $\mbi,~ \mbi^{\leqslant 0},~\mbi^{\geqslant 0},~$ and $\mbi^{>0}$ refer to the sets of integers, non-positive integers, non-negative integers, and positive integers, respectively.  NB:  $q \in \mathbb{Z}^{\geqslant 0}$.  An \& separates two linearly independent solutions.  DNO means that the case Does Not Occur. $\blsq$~means that one of the solutions contains a $\ln z$ term.  $\bltr$~means that one of the solutions is not one of the standard solutions.  $M$ is given by equation \eqref{Eq:M}, and $\wta$~is given by equations \eqref{Eq:M} and \eqref{Eq:wtM}. $U$~is given by equation \eqref{Eq:U1} and DLMF 13.2.42 when $b \not\in \mathbb{Z}$ and otherwise by the numbers in parentheses which refer to DLMF.}
	\label{Table:Labyrinth}
     \begin{center}
\begin{tabular}{||l|l|l|l|l||}
\hline \hline
   &COLUMN           &\;A       &\;B                    &\;C    \\ \hline
ROW&\quad$a$\textbackslash$b$&$b\not\in\mbi$&$b\in\mbi^{\leqslant0}$&$b\in\mbi^{>0}$\\\hline\hline
1  &$a\not\in\mbi$   &$M\&U$    &$\wta\&U$ (13.2.11)    &$M\&U$ (13.2.9),\\                            &$a-b\ne-(1+q)$   &$M\&\wta$	&and (13.2.9), or       &or (13.2.27)    \\
   &                 &$\wta\&U$ &(13.2.30)\qquad$\blsq$ &\qquad\qquad\quad$\blsq$\\ \hline
2  &$a\not\in \mbi$  &$M\&U$    &DNO                    &DNO    \\ 
   &$a-b=-(1+q)$     &$M\&\wta$ &                       &       \\  \hline
3  &$a\in\mbi^{\leqslant0}$&$M\&\wta$ &$\wta\&U$ (13.2.7),    &DNO   \\                     
   &$a-b\ne-(1+q)$   &$\wta\&U$ &or (13.2.32)           &      \\  \hline
4  &$a\in\mbi^{\leqslant0}$&DNO       &$\wta$ or $U$ (13.2.7),&$M$ or $U$ (13.2.7) \\    
   &$a-b=-(1+q)$     &          &or (13.2.8)            &or (13.2.10)  \\ 
   &                 &          &\,2nd sol. (13.2.31)   &\,2nd sol. (13.2.28) \\ 
   &                 &          & \qquad\qquad\quad$\blsq\;\bltr$& \qquad\qquad $\blsq\;\bltr$\\   \hline       
5  &$a\in\mbi^{>0}$  &$M\&U$    &$\wta\&U$ (13.2.11)    &$M\&U$ (13.2.9)     \\                     
   &$a-b\ne-(1+q)$   &$M\&\wta$ &and (13.2.9), or       &or (13.2.27)\\       
   &                 &$\wta\&U$ &(13.2.30)\qquad$\blsq$ &\qquad\qquad\quad$\blsq$\\  \hline  
6  &$a\in\mbi^{>0}$  &DNO       &DNO                    &$M\&U$ (13.2.9),  \\                     
   & $a-b=-(1+q)$    &          &                       &or (13.2.29) \\ \hline \hline
\end{tabular}
\end{center}  

\end{table} 

We also note that of the 12 twelve distinct cases that occur, not counting the DNO (do not occur) cases, only two, indicated by $\bltr$ in table~\ref{Table:Labyrinth}, require solutions that are not one of the three standard solutions, $M(a,b,z)$, $\wt$, and $U(a,b,z)$.  We find it interesting, and perhaps curious, that these are the two cases that give rise to the standard results for the bound states of the hydrogenic atom, although it is not the non-standard solutions that are relevant.

In addition, in nine of the 12 cases that occur, there is more than one way of choosing two linearly independent solutions (all of this is included in the table except for the fact that when $b=1, M = \wta$): 
\begin{enumerate}
\item Case 1.A.  $M, \widetilde{M}$, and $U$ are all valid, and so we can use any two of them.
\item Case 1.C.  For $b=1, \widetilde{M} = M$, and we can use $M$ and $U$ or $\widetilde{M}$ and $U$; otherwise, $\widetilde{M}$ is undefined and we must use $M$ and $U$. 
\item Case 2.A.  $U \propto \widetilde{M}$, and so we can use either $M$ and $U$ or $M$ and $\widetilde{M}$.
\item Case 3.A.  $U \propto M$, and so we can use either $M$ and $\widetilde{M}$ or $U$ and $\widetilde{M}$.
\item Case 4.B.  $M$ is not defined and $U \propto \widetilde{M}$, so that we can use either $\widetilde{M}$ or $U$ plus a non-standard second solution, given by 13.2.31 of DLMF.
\item Case 4.C.  $U \propto M, \widetilde{M} = M$\ for $b=1$, and $\widetilde{M}$ is not defined for $b \geqslant 2$; it follows that for $b = 1$ we can use any one of $M, \widetilde{M}$, and $U$; for $b \geqslant 2$, we can use either $M$ or $U$; in both cases, we also require a non-standard second solution, given by 13.2.28 of DLMF.
\item Case 5.A.  $M, \widetilde{M}$, and $U$ are all valid, and so we can use any two of them.
\item Case 5.C.  $\widetilde{M} = M$ for $b=1$ and $\widetilde{M}$ is not defined for $b\geqslant2$; so we can use either $M$ or $\widetilde{M}$ plus $U$ for $b=1$; we must use $M$ and $U$ for $b \geqslant 2$.
\item Case 6.C.  $\widetilde{M} = M$ for $b=1$ and $\widetilde{M}$ is not defined for $b\geqslant2$; so we can use either $M$ or $\widetilde{M}$ plus $U$ for $b=1$; we must use $M$ and $U$ for $b \geqslant 2$.
\end{enumerate}

Finally, we note that 13.2.27--13.2.32 of DLMF  yield, respectively, $U(a,b,z)$ for Case 1.C, the second solution for Case 4.C, $U(a,b,z)$ for Case 6.C, $U(a,b,z)$ for Case 1.B and Case 5.B, the second solution  for Case 4.B, and $U(a,b,z)$ for Case 3.B, aside from multiplicative constants. \\

The preferred way to use table~\ref{Table:Labyrinth} for a given problem is straightforward:
\begin{enumerate}
\item Based on the relevant values of $a$ and $b$, determine what cases can apply for the problem.
\item For each case, investigate whether the possible solutions satisfy the relevant boundary conditions.
\end{enumerate}

\section{Identities and limits}\label{S:Section4}

In section \ref{S:Basics}, we have discussed the definitions and basic properties of the \chf. In section \ref{S:Section3}, we have investigated the labyrinth of values of $a$ and $b$ and subjugated it to produce table~\ref{Table:Labyrinth}, which guides us in the choice of the solutions of Kummer's equation.  There are three additional topics necessary for the effective use of \chf. 

The first topic is the wealth of identities involving just \chf\; and the equally large set of identities involving \chf\;and their derivatives.  These are presented clearly in \S 13.3~(i) and \S 13.3~(ii), respectively of the DLMF, and in \S 13.4 of AS.

The second topic is the identification of the \chf\;with the various special functions.  This is done clearly and completely in \S 13.6 of DLMF  and \S 13.6 of AS \big(except that 13.6.2 and 13.6.5 of AS  are, of course, wrong if $b \in \mathbb{Z}^{\leqslant0}$\big).

The final topic that is absolutely necessary for the effective use of \chf\, in determining whether a putative solution satisfies the correct boundary conditions, is the limiting values of the \chf.

Throughout the discussions of $M(a,b,z)$ that follow, $b \not\in \mathbb{Z}^{\leqslant0}$, since $M(a,b,z)$ is not defined for $b \in \mathbb{Z}^{\leqslant0}$.

From equation  \eqref{Eq:M} it follows that
\begin{align} \label{Eq:limM1}
M(a,b,z) \sim &\frac{\gga (b)}{\gga (a)} \re^z z^{a-b} \left[1+\mathcal{O}(|z|^{-1})\right], \quad \text{as}\;z \to \infty, \quad \text{for}\;|\arg(z)| < \frac{\piup}{2} ,\;  
\text{for}\;a \not\in \mathbb{Z}^{\leqslant0}, 
\end{align}
that is, for $\Re z > 0$ and $a \not\in \mathbb{Z}^{\leqslant0}$.  (See 13.2.4 and 13.2.23 of DLMF, and 13.1.4 of AS.  Recall that $\arg(z)$ refers to the \textit{phase} of the generally complex number, $z$, and that DLMF  uses ph instead of $\arg$.)  More generally, \begin{align} \label{Eq:limM2}
M(a,b,z) \sim &\left[ \frac{\gga (b)}{\gga (a)} \re^z z^{a-b} + \frac{\gga(b)}{\gga(b-a)}\re^{\pm \ri \piup a}z^{-a}\right] \left[1+\mathcal{O}\left(z^{-1}\right)\right], \quad \text{as}\;z \to \infty,\; \\  &\text{for}\;- \frac{\piup}{2} < \pm  \arg(z) < \frac{3 \piup}{2} , \quad \text{unless}\;a \in \mathbb{Z}^{\leqslant0} \quad  \text{and}\;b-a \in \mathbb{Z}^{\leqslant0}. \notag
\end{align}
(See 13.2.4 and 13.7.2 of DLMF.)  The requirements on $\arg(z)$ correspond to branch cuts on the negative imaginary axis and on the positive imaginary axis, respectively.  (This is mostly, but not exactly, as given in 13.1.4, 13.1.5, and 13.5.1 of AS, and on page 60 of Slater~\cite{Slater}.) We explicitly state equation \eqref{Eq:limM1} even though it is included in equation \eqref{Eq:limM2} because it is the form most often needed.  In addition, for $a \in \mathbb{Z}^{\leqslant0}$, or equivalently for $a = - m$ with $m \in \mathbb{Z}^{\geqslant0}$,
\beq
M(-m,b,z) \sim z^m, \quad \text{for} \; z \to \infty ,
\eeq 
since $M(-m,b,z)$ is a polynomial in $z$ of $m$-th degree (see 13.2.7 of DLMF and the second entry in the table of 13.1.3 of AS).  More-or-less in connection with equation \eqref{Eq:limM2}, we note the first of the Kummer transformations,
\beq
M(a,b,z) = \re^z \; M(b-a,b,-z).
\eeq
(See 13.2.39 of DLMF and 13.1.27 of AS.)

For the limiting behavior of $U(a,b,z)$ as $z \to \infty$, we have
\beq
U(a,b,z) \sim z^{-a}\;\Big[1 + \mathcal{O}\left(z^{-1}\right)\Big], \quad \text{for}\;|\text{arg}(z)| < \;\piup .
\eeq
(See the comment after equation \eqref{Eq:U2}, 13.2.6 of DLMF, and 13.1.8 of AS.)

From equation  \eqref{Eq:M}, it is obvious that
\beq
M(a,b,z) \to 1, \quad \text{as}\;z \to 0.
\eeq
(See 13.2.13 of DLMF and 13.5.5 of AS.)

The limiting behavior of $U(a,b,z)$ as $z \to 0$ is rather more complicated than that of $M(a,b,z)$.  The simple part is 
\beq \label{Eq:limU1}
U(-m,b,z) = (-1)^m (b)_m + \mathcal{O}(z), \quad \text{as}\;z \to 0,
\eeq
and
\beq \label{Eq:limU2}
U(-q+b-1,b,z) = (-1)^q (2-b)_q \; z^{1-b} + \mathcal{O} \left(z^{2-b} \right), \quad \text{as}\;z \to 0,
\eeq
where $m \in \mathbb{Z}^{\geqslant 0}$ and $q \in \mathbb{Z}^{\geqslant 0}$ (see 13.2.14 and 13.2.15 of DLMF).  In all other cases:
\begin{IEEEeqnarray}{rCl}
 \text{a.}\quad U(a,b,z) &=& \frac{\Gamma(b-1)}{\Gamma(a)} z^{1-b} + \mathcal{O}\left(z^{2-\Re b} \right),\quad  \text{as}\;z \to 0, \quad \text{for}\;\Re b \geqslant 2, \, b \ne 2. \label{Eq:limU3} \\ 
\text{b.}\quad U(a,2,z) &=& \frac{1}{\gga(a)} z^{-1} + \mathcal{O}(\ln z), \quad \text{as}\;z \to 0. \label{Eq:limU4}\\ 
\text{c.}\quad U(a,b,z) &=& \frac{\gga(b-1)}{\gga(a)} z^{1-b} + \frac{\gga(1-b)}{\gga(1+a-b)} + \mathcal{O}\left(z^{2-\Re b}\right),\quad \text{as}\;z \to 0, \\
&&\text{for}\;1 \leqslant \Re b < 2,\, b \ne 1 . \notag \label{Eq:limU5} \\
\text{d.}\quad U(a,1,z) &=& - \frac{1}{\gga(a)} \left[\ln z + \psi (a) + 2 \gamma \right] + \mathcal{O}(z\ln z), \quad \text{as}\;z \to 0 , \\
&&\text{where}\;\psi (x) = {\gga ' (x)}/{\gga (x)}\;\text{is the digamma function and} \;\gamma\;\text{is Euler's constant}. \label{Eq:limU6} \nonumber  \\ 
\text{e.}\quad U(a,b,z) &=& \frac{\gga (1-b)}{\gga (1+a-b)} + \mathcal{O}\left(z^{1-\Re b}\right),\quad \text{as}\;z \to 0, \quad \text{for}\;0 < \Re (b) < 1 . \label{Eq:limU7} \\
\text{f.}\quad U(a,0,z) &=& \frac{1}{\gga (1+a)} + \mathcal{O}(z \ln z), \quad \text{as}\;z \to 0 . \label{Eq:limU8} \\
\text{g.}\quad U(a,b,z) &=& \frac{\gga (1-b)}{\gga (1+a-b)} + \mathcal{O}(z),\;\text{as}\;z \to 0\;, \quad \text{for}\;\Re b \leqslant 0,\, b \ne 0 . \label{Eq:limU9}
\end{IEEEeqnarray} 
\\
(See 13.2.16--13.2.22 of DLMF  and, except for the third equation above, 13.5.6 --13.5.12 of AS.)

\section{Application to the bound states of the hydrogenic atom} \label{S:Section5}

As an example for which solutions with $a \in \mathbb{Z}^{\leqslant0}$ and $b \in \mathbb{Z}$ are relevant, we consider the quantum-mechanical treatment of the bound states of the hydrogenic atom.
We know that the electron wavefunction has the usual separation of variables form for problems with spherical symmetry:
\beq
\psi_{\ell,m} (r,\theta,\phi) = R_{\ell} (r)\;Y_{\ell,m} (\theta,\phi),
\eeq
where $(r,\theta,\phi)$ are the standard spherical coordinates and $Y_{\ell,m} (\theta,\phi)$ denotes the usual spherical harmonics.  The radial wavefunction, $R_{\ell} (r)$, satisfies the second order, linear, ordinary differential equation,
\beq
\frac{\ud^2 \chi_{\ell} (r)}{\ud r^2} + \frac{2M}{\hbar^2}\Bigg[E + \frac{Z e^2}{4\piup\epsilon_0r} - \frac{\ell(\ell + 1)\hbar^2}{2Mr^2}\Bigg] \chi_{\ell} (r) = 0,
\eeq
where
\beq
\chi_{\ell} (r) \propto r\;R_{\ell} (r).
\eeq
Here, $M$ is the reduced mass of the electron, $\hbar$ is Planck's constant, $E$ is the internal energy of the atom, $Z$ is the atomic number of the nucleus, and we use SI units.

We take
\beq
z = ckr,
\eeq
where $c$ is a pure number and $k$ is a wavenumber.  For bound states, we take
\beq \label{Eq:Evsk}
E = - \frac{\hbar^2 k^2}{2M}.
\eeq
We thus obtain
\beq \label{Eq:Radial Eq}
\frac{\ud^2 \chi_{\ell} (r)}{\ud z^2} + \Bigg[- \frac{1}{c^2} + \frac{\gamma_c}{z} - \frac{\ell(\ell + 1)}{z^2}\Bigg] \chi_{\ell} (r) = 0,
\eeq
where
\beq
\gamma_c = \frac{2Z}{ck\tilde{a}_0}
\eeq
records the strength of the Coulomb interaction, and
\beq
\tilde{a}_0 = \frac{4\piup\epsilon_0\hbar^2}{Me^2}\;
\eeq
is the reduced Bohr radius (which uses the reduced mass of the electron).  We readily find that
\beq
\chi_{\ell} (r) \sim \re^{\pm z/c},\quad \text{as}\;z \to \infty,
\eeq
and
\beq
\chi_{\ell} (r) \sim z^{\ga}, \quad \text{with}\;\ga = \ell + 1 \quad \text{or}\;- \ell, \quad \text{as}\;z \to 0.
\eeq
Thus, without loss of generality, we take
\beq \label{Eq:chi}
\chi_{\ell} (r) = \re^{\pm z/c}\;z^{\ga}\;w_{\ell} (z), \quad \text{with}\;\ga =  \ell + 1 \quad \text{or}\;- \ell.
\eeq
Upon substituting this into equation~\eqref{Eq:Radial Eq}, we readily obtain
\beq
z\frac{\ud^2 w_{\ell}}{\ud z^2} + 2 \left( \ga \pm \frac{z}{c}\right) \frac{\ud w_{\ell}}{\ud z} + \left( \gamma_c \pm 2 \frac{\ga}{c}\right) w_{\ell} = 0.
\label{Eq:chf_hydrogen}
\eeq
We choose the ``$-$'' sign and $c = 2$, so that this reduces to the \che, in equation~\eqref{Eq:che1}, with
\beq
a = \frac{b}{2} - \frac{Z}{k\tilde{a}_0} \quad \text{and}\;b = 2(\ell + 1) \quad \text{or}\;b=-2\ell.
\label{Eq:a_b_parameters}
\eeq
Then we have
\beq \label{Eq:chi_l}
\chi_{\ell} (r) = e^{-z/2}\;z^{b/2}\;w_{\ell} (z), \quad \text{with}\;b=2(\ell+1) \quad \text{or}\;-2\ell.
\eeq

Although it may be tempting to reject $ b = - 2 \ell$ on the grounds that it indicates that $\chi_{\ell} \not\to 0$ as $z \to 0$, we choose not to do so, at least not until we see how $w_{\ell}(z)$ behaves as $z \to 0$.  Equation \eqref{Eq:che1} is a second-order, linear, ordinary differential equation, which thus has two linearly independent solutions, and we must use the boundary conditions to determine the proper solutions to solve the problem. A typical strategy, that is taught in quantum mechanics classes and employed in quantum mechanics textbooks, is to pick $M$ and~$U$ as the linearly independent solutions and then systematically eliminate solutions that fail to satisfy the boundary conditions. However, this approach is fundamentally flawed in that $M$ and $U$ are not always linearly independent solutions, because sometimes $U \propto M$, and even worse, we sometimes are faced with values of $b$ such that $M$ is not even defined. As we have seen in the analysis leading to table~\ref{Table:Labyrinth}, the proper starting position is to consider $M, U$, and $\widetilde{M}$ as possible solutions, note that \textit{a priori} any two of them may be linearly independent solutions, and use the values of $a$ and $b$ to determine if we can use two of them as the linearly independent solutions, and if so, which two.  In this process, we also learn that it is sometimes necessary to use yet another function as a linearly independent solution. Hence, one must proceed very carefully, avoiding errors in logic, in order to solve the problem in full generality.

As indicated, we are allowed two linearly independent solutions of Kummer's equation. The immediate issue is how to label the solutions.  The first label is of course $\ell$, the orbital angular momentum quantum number.  Since there are two possible sets of values for $a$ and $b$ in equation \eqref{Eq:chf_hydrogen} (with the ``$-$'' sign and $c = 2$), as indicated in equation \eqref{Eq:a_b_parameters}, we use a second subscript, $v$, with values $1$ and $2$, to denote these two different choices for the parameters $a$ and $b$, which correspond to different entries in table~\ref{Table:Labyrinth}.  We then use a third and final subscript, which also takes on the values 1 and 2, to denote the two linearly independent solutions.  Accordingly, we write our general solution, for given values of $\ell$ and $v$, and showing all the labels and arguments, as
\beq
\chi_{\ell v} (a_v , b_v , z) = \re^{-z/2} z^{b_v /2} \Big[\mathcal{C}_{\ell v 1} w_{\ell v1}(a_v , b_v ,z) + \mathcal{C}_{\ell v 2} w_{\ell v2}(a_v , b_v ,z)\Big], \quad v=1\;\text{or}\;2,
\eeq
where $\mathcal{C}_{\ell v 1}$ and $\mathcal{C}_{\ell v 2}$ are constants,
\beq
a_1 = \ell + 1 - \frac{Z}{k\tilde{a}_0},  \quad b_1 = 2(\ell + 1)  \quad \text{and}\;a_2 = - \ell - \frac{Z}{k\tilde{a}_0}, \quad  b_2 = - 2 \ell ,
\eeq
and  $w_{\ell v1}(a_v , b_v ,z)$ and $w_{\ell v2}(a_v , b_v ,z)$ are the two linearly independent solutions for the given values of $\ell$ and $v$. We are very careful to note that we have two linearly independent solutions here, as there is no general notation we can use to explicitly specify them \textit{a priori}, due to the issues that we discussed above. Instead, we need to refer to table~\ref{Table:Labyrinth} in making our way through the labyrinth of values of $a$ and $b$.

We first consider $v = 1$.  Since $b_1 = 2(\ell+1)$ is a positive integer, we are required to consider cases 1.C, 4.C, 5.C, and 6.C.  Since $M(a,b,z)$ and $U(a,b,z)$ are solutions for all of these cases, except that they are not distinct solutions for Case 4.C, it makes sense that we take
\beq
w_{\ell 11}(a_1 , b_1 , z) = M(a_1 , b_1, z).
\eeq
We know from equation \eqref{Eq:limM1}, and 13.2.4 and either 13.2.23 or 13.7.2 of DLMF, or 13.1.4 of AS, that for $a \not\in \mathbb{Z}^{\leqslant0}$,
\beq
M(a,b,z) \sim \frac{\Gamma(b)}{\Gamma(a)} z^{a-b}\re^z, \quad \text{as}\;z \to \infty,
\eeq 
so that
\beq
e^{-z/2}z^{b_1/2}w_{\ell 11}(a_1,b_1,z) \sim z^{-Z/k\tilde{a}_0}\re^{z/2}, \quad \text{as}\;z \to \infty.
\eeq
This means we cannot satisfy the requirement of a normalizable and everywhere finite wavefunction when $a_1\not\in \mathbb{Z}^{\leqslant0}$.
Hence, we must have $a_1 \in \mathbb{Z}^{\leqslant0}$.  Accordingly, we choose
\beq \label{Eq:n_r}
a_1 = \ell + 1 - \frac{Z}{k\tilde{a}_0} = - n_r,
\eeq 
where $n_r \in \mathbb{Z}^{\geqslant0}$.  This immediately limits us to Case 4.C.  The second solution then is not given by $U(a_1,b_1,z)$ or $\widetilde{M}(a_1,b_1,z)$, but rather by 13.2.28 of DLMF. However, since $(a)_0 = (-n_r)_0 = 1$, 13.2.28 of DLMF always has a $\ln z$ term, and so the second solution of Case 4.C is unacceptable, which consequently requires $\mathcal{C}_{\ell 12} = 0$.  We next define the principal quantum number,
\beq \label{Eq:n}
n = n_r + \ell + 1,
\eeq 
and note that since $n_r \in \mathbb{Z}^{\geqslant0},\, n \geqslant \ell + 1$.  Equations \eqref{Eq:n_r} and \eqref{Eq:n} with $k \to k_n$ yield
\beq \label{Eq:kn}
\frac{Z}{k_n \tilde{a}_0} = n.
\eeq
Consequently, from equations \eqref{Eq:Evsk} and \eqref{Eq:kn}, we obtain
\beq \label{Eq:Energy}
E_n = - \frac{1}{n^2}\;\frac{Z^2 e^2}{8\piup\epsilon_0 \tilde{a}_0}.
\eeq
This is, of course, the usual result for the bound state energies of the hydrogenic atom.  The radial wavefunctions are then given by
\beq
R_{n\ell}(r) \propto \re^{-k_n r} (k_n r)^{\ell} M(-n+\ell+1,2\ell+2,2k_n r).
\eeq
According to 13.6.19 of DLMF  and 13.6.9 of AS,
\beq
M(-n+\ell+1,2\ell+2,2k_n r) \propto L_{n-\ell-1}^{(2\ell+1)} (2k_n r),
\eeq
where the $L_{n-\ell-1}^{(2\ell+1)}$ are the associated Laguerre functions.  Consequently,
\beq \label{Eq:wavefcn}
R_{n\ell} (r) = \mathcal{N}_{n\ell} \re^{-k_n r} (k_n r)^{\ell} L_{n-\ell-1}^{(2\ell+1)} (2k_n r),
\eeq
where the $\mathcal{N}_{n\ell}$ are the normalization constants.  Equations \eqref{Eq:Energy} and \eqref{Eq:wavefcn} are the usual results.

We note that instead we could have  used $U(a_1,b_1,z)$ as our first attempt at a solution.  In that case, it is the behavior of $U(a_1,b_1,z)$ as $r \to 0$, as given by equations \eqref{Eq:limU3} and \eqref{Eq:limU4}, that forces us to take $a_1 \in \mbi^{\leqslant0}$ and restrict our considerations to Case 4.C.  This yields the usual spectrum, and since $U(a,b,z) \propto M(a,b,z)$ when $a \in \mbi^{\leqslant0}$ and $b \not\in \mbi^{\leqslant0}$, the usual wavefunctions also follow with the same argument as given above.

Usually, a physicist would stop at this point, since a solution that satisfies all the requirements has been obtained.  However, it is instructive to consider the remaining possibilities. The obvious remaining possibility is the solution for $v = 2$.  However, we cannot be assured that we have even finished with the solution for $v = 1$.  For Case 1.C, 5.C, and 6.C, $M(a_1,b_1,z)$ cannot be prevented from diverging as $r \to \infty$, and for Case 1.C and Case 5.C, $U(a_1,b_1,z)$ contains $\ln z$ terms.  However, the second solution for Case 6.C is potentially acceptable and so we really should consider it.

Since $a_1 \not\in \mathbb{Z}^{\leqslant0}$ for Case 6.C, $\mathcal{C}_{\ell11} = 0$ and
\beq
w_{\ell12}(a_1,b_1,z) = U(a_1,b_1,z).
\eeq
Since $a \in \mbi^{> 0}$ for Case 6.C, we take $\ell - \tfrac{Z}{k \tilde{a}_0} = m$, where $m \in \mathbb{Z}^{\geqslant0}$, which gives $a_1 = 1+m$~.  Then
\beq
w_{\ell12}(a_1,b_1,z) = U(1+m,2\ell +2,z).
\eeq
According to equations \eqref{Eq:limU2}--\eqref{Eq:limU4}, and DLMF 13.2.15--13.2.17, or AS 13.5.6 and 13.5.7,
\beq
w_{\ell 12}(1+m,2\ell +2,z) \sim z^{-1-2\ell} \quad \text{as}\;z \to 0,
\eeq
so that
\beq
\chi_{\ell 2}(r) \sim \re^{-z/2}\;z^{\ell +1}\;z^{-1-2\ell} = \re^{-z/2}\;z^{-\ell} \not\to 0 \quad \text{as}\;z \to 0.
\eeq
Consequently, this case does not yield an acceptable solution.

We thus turn to $v = 2$ and the second set of values, $a_2$ and $b_2$.  We have $b_2 = - 2\ell$, which $ \in \mathbb{Z}^{\leqslant0}$.  Thus, we must consider cases 1.B, 3.B, 4.B, and 5.B.  Since for all of these cases, both $\widetilde{M}(a_2,b_2,z)$ and $U(a_2,b_2,z)$ are solutions, but not distinct solutions for Case 4.B, it makes sense to take
\beq
w_{\ell 21} (a_2 , b_2 , z) = \widetilde{M}(a_2,b_2,z),
\eeq
or, more explicitly, with the use of equation ~\eqref{Eq:wtM}, 
\beq \label{Eq:v=2 first sol}
w_{\ell 21} (a_2 , b_2 , z) = z^{1-b_2} M(1+a_2 -b_2 , 2-b_2 , z).
\eeq
The reason that we should not reject $\chi_{\ell} \propto z^{\ga}$ with $\ga = - \ell$ as $z \to 0$ now becomes 
clear.  With equations~\eqref{Eq:chi_l} and~\eqref{Eq:v=2 first sol}, $\chi_{\ell} \sim z^{b_2/2} z^{1-b_2} = 
z^{\ell +1}$ as $z \to 0$.  Thus, it would have been wrong to reject the asymptotic behavior $\chi_{\ell} \sim z^{\ga}$ with $\ga = - \ell$ as $z \to 0$. 

According to 13.2.4 and either 13.2.23 or 13.7.2 of DLMF, we have, for $1+a_2 - b_2 \not\in \mathbb{Z}^{\leqslant0}$,
\beq
w_{\ell 21} (a_2 , b_2 , z) \sim \frac{\Gamma(2-b_2)}{\Gamma(1+a_2-b_2)} z^{1 + a_2 - 2 b_2 } \re^z, \quad \text{as}\;z \to \infty , 
\eeq
so that, for $1+a_2 - b_2 \not\in \mathbb{Z}^{\leqslant0}$,
\beq
\re^{-z/2} z^{b_2/2}\;w_{\ell21}(a_2,b_2,z) \sim z^{1 + a_2 - 3 b_2/2} \re^{z/2},\quad \text{as}\;z \to \infty.
\eeq
So the only way that we can have a wavefunction that is normalizable and everywhere finite is for the excluded case,  $1+a_2 - b_2 \in \mathbb{Z}^{\leqslant0}$.  Of course,
\beq
1 + a_2 - b_2 = 1 - \ell - \frac{Z}{k \tilde{a}_0} + 2\ell =  \ell +1 - \frac{Z}{k \tilde{a}_0}.   
\eeq
So we can take $1 + a_2 - b_2 = - n_r $, with $n_r \in \mathbb{Z}^{\leqslant0}$, which implies the energy quantization condition given in equation~\eqref{Eq:Energy}.  Then we have
\beq
M(1+a_2 -b_2,2-b_2 ,z) = M(-n_r , 2\ell + 2,z),
\eeq
which also yields equation  \eqref{Eq:wavefcn}.  We note that
\beq
1 + a_2 - b_2 = - n_r \Longrightarrow a_2 = - 1 - 2\ell - n_r,
\eeq
showing that $a_2 \in \mathbb{Z}^{\leqslant 0}$. Furthermore,
\beq
a_2 - b_2 = - (1 + n_r ),
\eeq
These two constraints dictate Case 4.B.  The second solution is given by 13.2.31 of DLMF with $a = a_2$ and $n = 2\ell$.  Then $a+n+1 = -n_r$, and since $(-n_r)_0 =1$, this solution always has a $\ln z$ term and must be rejected.  Thus the second choice of parameters, $a_2$ and $b_2$, yields the same result as the more commonly used first choice.

We note that just as for $v = 1$, we could have instead used $U(a_2,b_2,z)$ as our first attempt at a solution.  In that case, it is again the behavior of $U(a_2,b_2,z)$ as $r \to 0$ that forces us to take $a_2 \in \mbi^{\leqslant0}$ and restrict our considerations to Case 4.B.  This yields the usual spectrum, and since $U(a,b,z) \propto M(a,b,z)$ when $a \in \mbi^{\leqslant0}$ and $b \not\in \mbi^{\leqslant0}$, the usual wavefunctions again follow.

Just as there was a second possibility for $v = 1$, so there is a second possibility for $v = 2$.  Since $1+a_2 -b_2 \not\in \mathbb{Z}^{\leqslant0}$ for Case 1.B, Case 3.B, and Case 5.B, $\widetilde{M}$ cannot be prevented from diverging for these cases; moreover, the second solutions for Case 1.B and 5.B contain $\ln z$ terms.  What remains is the second solution of Case 3.B.  Thus, we take $\mathcal{C}_{\ell21} = 0$ and
\beq
w_{\ell22}(a_2,b_2,z) = U(a_2,b_2,z).
\eeq
We take $a_2 = -m$, where $m \in \mathbb{Z}^{\geqslant 0}$.  Then,
\beq
w_{\ell 22}(a_2,b_2,z) = U(-m,-2\ell,z).
\eeq
According to equations \eqref{Eq:limU1}, or DLMF 13.2.14,
\beq
w_{\ell 22}(a_2,b_2,z) = (-1)^m (-2\ell)_m + \mathcal{O}(z).
\eeq
The requirement $a_2 - b_2 \ne -(1+q)$, where $q \in \mathbb{Z}^{\geqslant 0}$,\; $\Longrightarrow -m + 2\ell \geqslant 0,\, -2\ell + m \leqslant 0$, and $-2\ell +m -1 \leqslant -1$, so that $(-2\ell)_m \ne 0$.  Then,
 \beq
 \chi_{\ell 2}(r) \sim \re^{-z/2}\;z^{-\ell} \not\to 0 \quad \text{as}\;z \to 0.
 \eeq
Thus, Case 3.B cannot yield an acceptable solution. 

We emphasize that one should exhaust  all of the possibilities in the table for a given value of $b$, including looking at the first and second solutions, before moving on to the next value of $b$.

This completes the analysis for the solution of the hydrogenic atom problem in quantum mechanics. Note that the two choices for $b$, namely $2(\ell + 1)$ and $-2\ell$, result in the same energy spectrum and the same wavefunctions.   In other words, there is no basis for rejecting $\ga = - \ell$ in equation \eqref{Eq:chi}.  This is a fact that is not noted in most, perhaps all, quantum mechanics textbooks.  

It is remarkable and curious that the two cases that yield the standard results, Case 4.C for $b = 2(\ell + 1)$ and Case 4.B for $b = - 2\ell$, are the two cases where the second linearly independent solution is not one of $M(a,b,z), \wt$, or $U(a,b,z)$.

Note how systematically and smoothly, and even spectacularly, table~\ref{Table:Labyrinth} guides and facilitates this analysis.

In addition, the use of the confluent hypergeometric functions to explain the Rydberg series goes further than just explaining the spectrum of hydrogen. In a series of four seminal papers~\cite{Hartree1,Hartree2,Hartree3,Hartree4}, Hartree worked out how the systematics of the Rydberg series for other atoms could be understood quantitatively in terms of the properties of confluent hypergeometric functions.  This work led to the origin of quantum defect theory~\cite{Aymar}, in which the integer that appears in the formula for the hydrogen spectrum is replaced for the alkalis by a fractionally shifted integer, something which had already been observed in experimental data in the 1920s.

\section{The cutoff Coulomb potential} \label{S:Section6}

As an interesting and instructive example of the care that is necessary in working with \chf, let us consider the cutoff Coulomb potential discussed by Othman, de Montigny, and Marsiglio~\cite{Othman et al.}. The potential is given by
\beq
V(r) = 
\begin{cases}
- {e^2}/{4\piup\epsilon_0r_0}, \quad \text{for $0 \leqslant r \leqslant r_0$, region I,}\\ 
-{e^2}/{4\piup\epsilon_0r}, \quad  \text{for $r \geqslant r_0$, region II,}
\end{cases}
\eeq
where we again use SI units.  The authors take
\beq \label{Eq:u form 1}
\chi_{\ell}(\rho) = \rho^{\ell + 1} \re^{-\rho} v(\rho),
\eeq
where $\chi_{\ell}(\rho)$ is denoted as  $u(r)$ in reference~\cite{Othman et al.}, and $\rho = k r$.  They show that in region II, $v$ satisfies the \che\;with $a = \ell +1 - \rho_0/2$, where $\rho_0 = 2/k a_0$ with $a_0$ the Bohr radius, $b = 2(\ell + 1)$, and $z = 2\rho$.  They then restrict their considerations to $\ell = 0$, which simplifies, but does not detract from the subsequent analysis.   

The authors correctly note that $a$ must be reserved for matching the wavefunctions and their derivatives with respect to $r$ at $r_0$, and consequently one is not free to choose $a \in \mathbb{Z}^{\leqslant0}$ to prevent $M(a,b,z)$ from diverging as $z \to \infty$.  The authors assume that $a \not \in \mathbb{Z}^{\leqslant 0}$, and accordingly drop $M(a,b,z)$ and use $U(a,b,z)$ in region II.  They find that matching the wavefunctions and their derivatives with respect to $r$ at $r_0$ results in $a \not \in \mathbb{Z}$, which corresponds to Case 1.C in the table.  This indicates that the assumption that $a \not\in \mathbb{Z}^{\leqslant0}$ is correct.  

The authors then show that in the limit $r_0 \to 0$,  $\rho_0 \to 2n$, where $n$ is the principal quantum number, and consequently $a \to$ an element of $\mathbb{Z}^{\leqslant0}$, so that the usual energy spectrum follows.  Moreover, the Tricomi function is rendered finite at $r = 0$.   The authors then note that the Tricomi function is an associated Laguerre function, and thus the standard results for the bound state wavefunctions follow.  Since they are taking a limit, they always remain in Case 1.C. 

However, solving the problem at $r_0 = 0$ requires a different analysis.  One can continue to use an \textit{ansatz} that the solution is a linear combination of $M$ and $U$ for all cases except Case 4.C.  After discovering that none of those solutions satisfy the boundary conditions, one finds it necessary to use Case 4.C, for which the \textit{ansatz} is different.  Even though the solutions found for $r_0 =0$ and for $r_0 \to 0$ are exactly the same, the \textit{ansatz} and the procedure for obtaining them are different.  Curiously, as the authors take the limit as $r_0 \to 0$, they find that $a \to$ an element of $\mathbb{Z}^{\leqslant0}$, and $M(a,b,z)$ and $U(a,b,z)$ are the same aside from an irrelevant multiplicative factor that is independent of $z$.  Hence, there should be no surprise that the usual results for hydrogen follow when $r_0 \to 0$.

The authors conclude from their results that `` \ldots even when we consider the usual Coulomb potential without a cutoff we should include the Tricomi function as well as the Kummer solution.''.  However, even this is in general inadequate, in that initially one should consider $M(a,b,z)$, which is the Kummer function, the first power series solution of the \che, $U(a,b,z)$, the Tricomi function, and $\wt$, which is the second power series solution of the \che, as potential solutions.  Morever, as we discussed in the previous section, for the hydrogenic atom, even after dispensing with $\wt$, which is the same as $M(a,b,z)$ for $b=1$ and undefined for $b \in \mathbb{Z}^{\geqslant2}$, this initial approach indicates that cases 1.C, 5.C, and 6.C must be excluded.  One thus settles on the necessity of using Case 4.C, and finds that since the Tricomi function and the Kummer function are proportional to one another, it is necessary to find an additional solution that is not one of the three standard solutions for use in the \textit{ansatz}.  As a consequence, one is led to consider the solution given by 13.2.28 of DLMF.  Since that solution diverges at $r = 0$, it can be excluded.

Our final point is that most, perhaps all, discussions of hydrogen and closely related systems do not consider 
\beq \label{Eq:u form 2}
\chi_{\ell}(\rho) = \rho^{- \ell} \re^{-\rho} v(\rho).
\eeq
As we have seen in the previous section, both of equations \eqref{Eq:u form 1} and \eqref{Eq:u form 2} follow from the behavior of $\chi_{\ell}(\rho)$ as $r \to \infty$ and $r \to 0$.  Equation \eqref{Eq:u form 2} leads to the \che\;for $v$ with $a = - \ell - \rho_0/2$ and $b = - 2\ell$.  Much the same complications that occur when equation \eqref{Eq:u form 1} is used also occur when equation \eqref{Eq:u form 2} is used.  The main difference is that since $b \in \mathbb{Z}^{\leqslant0}$, it is $M(a,b,z)$ rather than $\wta$ that must be excluded at the start.  Moreover, equations \eqref{Eq:u form 1} and \eqref{Eq:u form 2} both lead to the standard bound state energies and wavefunctions.

All of these considerations are fully taken into account in table~\ref{Table:Labyrinth} and the analysis in the Appendix, which results in table~\ref{Table:Labyrinth}.  Accordingly, we see in a very clear and concrete way that when working with \chf, one must carefully note the character of $a$ and $b$ ($\not \in \mathbb{Z}$, $\in \mathbb{Z}^{\leqslant0}$, $\in \mathbb{Z}^{>0}$, $\in \mathbb{Z}^{\geqslant 2}$) and the constraint on $a - b$ [$a-b \ne -(1+q)$ or $a-b =-(1+q)$, where $q \in\mathbb{Z}^{\geqslant0}$] and choose the two linearly independent solutions accordingly.
 
\section{Summary and discussion} \label{S:Section7}

We have carefully and thoroughly discussed the definitions and basic properties of the \chf\;that are necessary to effectively employ them in the solution of many problems in physics, as well as in other areas of science, engineering, and even mathematics.  We presented the basic definitions of the Kummer and Tricomi functions, $M(a,b,z)$ and $U(a,b,z)$, respectively.  We noted that $M(a,b,z)$ and $\wt \equiv z^{1-b} M(1+a-b,2-b,z)$ are the two power series solutions of the \che\;and, together with $U(a,b,z)$, are the three standard solutions with which it is most convenient to begin considerations.  We discussed in detail the circumstances in which these three solutions are or are not defined, and are and are not distinct, emphasizing the complicated ways in which the characters of $a$ and $b$ and the constraints on $a-b$ are all-important.  We emphasized the great care that is needed in determining how these three standard solutions can be used to obtain two linearly independent solutions and the circumstances in which they cannot.  We noted that the numerous identities involving just these functions and those involving these functions and their derivatives are presented very clearly in the DLMF  and AS.   We also noted that the identification of the \chf\;with the various special functions is presented clearly and completely in DLMF  and, for the most part, in AS.  We also presented the limiting values as $z \to \infty$ and as $ z \to 0$ that are needed to ensure that the boundary conditions of the problem being considered are obeyed.

We believe that the most striking and useful result of our efforts is our navigation of the convoluted and complicated labyrinth of values of $a$ and $b$ in which we emphasized the determination of two linearly independent solutions of the \che, how to obtain $U(a,b,z)$ when it is one of the two linearly independent solutions, and what to do when only a single one of the three standard solutions is distinct or survives.  We present the details of this effort in the Appendix and the results of this effort in section~\ref{S:Section3} in the form of a very comprehensive table~\ref{Table:Labyrinth}, which we expect to be of considerable use in the employment of the \chf.   

We carefully apply all of this to what for many is a very familiar example, the problem of the quantum-mechanical bound states of the hydrogenic atom.  Our treatment of this problem is complete in that we do not reject the case where $\chi_{\ell} \sim z^{-\ell}$ as $r \to 0$.  That is, we consider both $b = 2{\ell + 1}$ and $b = - 2 \ell$, and we also consider the nonstandard second solutions in both cases.  We find that the extent to which table~\ref{Table:Labyrinth} facilitates the use of \chf\;in solving the Schr\"{o}dinger equation is startling.  We would expect this to be the case in other problems involving \chf. 

We also considered the discussion of the bound state energies and wavefunctions of the cutoff Coulomb potential considered by Othman, de Montigny, and Marsiglio~\cite{Othman et al.}.  We saw that this problem emphasizes that considerable care is necessary in working with \chf.

We have endeavored to prepare this guide so that it will substantially aid the instruction and research that involves Kummer's differential equation and the \chf, especially for energy eigenvalue problems in quantum mechanics, but also for other areas in science, engineering, and even mathematics.

\section*{Acknowledgements}

We would like to thank Frank Marsiglio and Ian J. Thompson for a careful reading of the manuscript and extensive discussion of section~\ref{S:Section6}. We also thank Charles W. Clark for pointing out the importance of the confluent hypergeometric functions in working out the details of quantum defect theory by Hartree and its application to the multichannel Rydberg spectroscopy of complex atoms. 
This work was supported by the National Science Foundation under grant number PHY-1915130. In addition, JKF was also supported by the McDevitt bequest at Georgetown University.

\appendix
\section*{Appendix:  Analysis of the labyrinth of values of $a$ and $b$}
\setcounter{section}{1}

What follows gives the reasoning that results in table~\ref{Table:Labyrinth}, which is in section \ref{S:Section3}.  We would expect that taking the time and effort to follow and understand this reasoning would facilitate the intelligent and efficient use of the table.  This is, at least in part, because  working through this Appendix should serve to sensitize the reader to the intricacies of \chf.  

We begin with the first category of $a$ and work through the three categories of $b$ that go with it.  We then proceed to the second category of $a$ and go through the three categories of $b$, etc.

For each case, i.e., for each set of values of $a$ and $b$, we proceed as follows:
\begin{enumerate}
\item If the constraint on $a-b$ indicates that the case does not occur, so note.
\item Note any further consequences of the constraint on $a-b$.
\item Note which of the standard solutions, $M(a,b,z), \wt$, and $U(a,b,z)$, survive.
\item If $U(a,b,z)$ is one of the surviving solutions, note how to determine it.
\item If only one of the three standard solutions survives, or if only two of the standard solutions survive, but they are not linearly independent, note how to obtain a second linearly independent solution.
\item For column A of the table, where multiple choices for the two linearly independent solutions are possible, rank order the possible choices.  For columns B and C, note the preferred choice for the two linearly independent solutions.
\end{enumerate}

Our criteria for determining the preferred choice for the two linearly independent solutions to use are:

\begin{enumerate}
\item If $M(a,b,z)$ and $U(a,b,z)$ are both defined and linearly independent of one another, we use them.
\item If $U(a,b,z) \propto M(a,b,z)$, we use $M(a,b,z)$.
\item If $U(a,b,z) \propto \wt$, we use $\wt$.
\item If $b = 1$, we use $M(a,b,z)$.
\end{enumerate}

\noindent \underline{Case 1.A.  $a \not\in \mathbb{Z}$ with $a-b \ne - (1+q)$, and  $b \not\in \mathbb{Z}$}

$a-b\ne-(1+q) \Longrightarrow$ either $a-b \not\in \mathbb{Z}$ , or $a\geqslant b$.  Thus, this case complements Case 2.A, for which~$a < b$.

All of $M(a,b,z),\;\wt$, and $U(a,b,z)$ are solutions and are linearly independent of one another.

$U(a,b,z)$ is given by equation \eqref{Eq:U1} or \eqref{Eq:U2}, or 13.2.42 of DLMF, or 13.1.3 of DLMF.

We can use any two of the solutions.   Most often today, one uses $M(a,b,z)$ and $U(a,b,z)$, but $M(a,b,z)$ and $\wt$ are also used, and of course $\wt$ and $U(a,b,z)$ could also be used.
\\

\noindent \underline{Case 1.B.  $a \not\in \mathbb{Z}$ with $a-b \ne - (1+q)$, and $b \in \mathbb{Z}^{\leqslant 0}$}

$a \not\in \mathbb{Z}$ and $b \in \mathbb{Z}^{\leqslant 0}$ ensures $a-b \ne - (1+q)$.

$b \in\mathbb{Z}^{\leqslant 0} \Longrightarrow M(a,b,z)$ is not defined; $a \not\in \mathbb{Z}$ and $b \in \mathbb{Z}^{\leqslant 0} \Longrightarrow \wta$ is defined and is linearly independent of $U(a,b,z)$, which is also defined.

$U(a,b,z)$ is given by 13.2.11 and 13.2.9 of DLMF, or by 13.2.30 of DLMF, or with care by 13.1.7 and 13.1.6 of AS.  Note that $U(a,b,z)$ contains $\ln z$ terms.

Thus, with $b = - n$, where $n \in \mathbb{Z}^{\geqslant 0}$, we use $\widetilde{M}(a,-n,z)$ and $U(a,-n,z)$.
\\
\\

\noindent \underline{Case 1.C.  $a \not\in\mathbb{Z} $ with $a-b \ne - (1+q)$, and $b \in \mathbb{Z}^{> 0}$}

$a \not\in \mathbb{Z}$ and $b \in \mathbb{Z}^{> 0}$ ensures $a-b \ne - (1+q)$.

$b \in \mathbb{Z}^{> 0} \Longrightarrow \wt = M(a,b,z)$ for $b = 1$, and $\wt$ is not defined for $b \in \mathbb{Z}^{\geqslant 2}$.  $M(a,b,z)$ and $U(a,b,z)$ are linearly independent solutions.

$U(a,b,z)$ is given by 13.2.9 and also 13.2.27 of DLMF, or by 13.1.6 of AS.  Note that it contains~$\ln z$ terms.

With $b = 1+n$, where $n \in \mathbb{Z}^{\geqslant 0}$, we take $M(a,1+n,z)$ and $U(a,1+n,z)$.
\\

\noindent \underline{Case 2.A.  $a \not\in \mathbb{Z}$ with $a-b = - (1+q)$, and $b \not\in \mathbb{Z}$}

$a-b = - (1+q) \Longrightarrow a < b$.  Thus, this case complements Case 1.A, for which $a \geqslant b$ if $a - b \in \mathbb{Z}$.

Since $b = 1+a+q$, according to 13.2.8 of DLMF, $U(a,b,z) \propto \wt$.

$U(a,b,z)$ is given by equation \eqref{Eq:U1} or 13.2.8 and 13.2.42 of DLMF.

We choose $M(a,b,z)$ and $U(a,b,z)$, with $b = 1+a+q$.  Of course, we could also use $M(a,b,z)$ and~$\wt$\;.
\\ 

\noindent \underline{Case 2.B.  $a \not\in\mathbb{Z}$ with $a-b = - (1+q), b \in \mathbb{Z}^{\leqslant 0}$}

This case does not occur, because $a \not\in \mathbb{Z}$ and $b \in \mathbb{Z}^{\leqslant 0}$ precludes $a-b = - (1+q)$.
\\

\noindent \underline{Case 2.C.  $a \not\in\mathbb{Z} $ with $a-b = - (1+q), b \in \mathbb{Z}^{> 0}$}

This case also does not occur, because $a \not\in \mathbb{Z}$ and $b \in \mathbb{Z}^{> 0}$ precludes $a-b = - (1+q)$.
\\

\noindent \underline{Case 3.A.  $a \in \mathbb{Z}^{\leqslant 0}$ with $a-b \ne - (1+q)$, and $b \not\in \mathbb{Z}$}

$a \in \mathbb{Z}^{\leqslant 0}$ and $b \not\in \mathbb{Z}$ ensures that $a-b \ne - (1+q)$.

$a \in \mathbb{Z}^{\leqslant 0}$ and $b \not\in \mathbb{Z}^{\leqslant 0} \Longrightarrow U(a,b,z) \propto M(a,b,z)$, according to 13.2.7 of DLMF.

$U(a,b,z)$ is given by 13.2.7 or 13.2.42 of DLMF.

With $a = - m$, where $m \in \mathbb{Z}^{\geqslant 0}$, we use $M(-m,b,z)$ and $\wta(-m,b,z)$.  We could also use $U(-m,b,z)$ and $\wta(-m,b,z)$. Of course, since $U(-m,b,z) \propto M(-m,b,z)$, the two choices are essentially the same.
\\

\noindent \underline{Case 3.B.  $a \in \mathbb{Z}^{\leqslant 0}$ with $a-b \ne - (1+q)$, and $b \in \mathbb{Z}^{\leqslant 0}$}

With $a = - m$ and $ b = - n$, where $m \in \mathbb{Z}^{\geqslant 0}$ and $n \in \mathbb{Z}^{\geqslant 0}$, the constraint on $a-b$ implies $m \leqslant n$.  Thus, this case complements Case 4.B.

We know that $b \in \mathbb{Z}^{\leqslant 0} \Longrightarrow M(a,b,z)$ is not defined and $\wt$ is defined.  Thus, $U(a,b,z)$ and~$\wt$ are linearly independent solutions.

Also, $a \in \mathbb{Z}^{\leqslant 0}, b \in \mathbb{Z}^{\leqslant 0} \Longrightarrow U(a,b,z)$ is given by  13.2.7 of DLMF, with the contents between the two ='s deleted, or by 13.2.32 of DLMF.  Note despite the fact that $b \in \mathbb{Z}, U(a,b,z)$ does not contain any~$\ln z$ terms.

It follows that we can use $\wta(-m,-n,z)$ and $U(-m,-n,z)$.
\\

\noindent \underline{Case 3.C.  $a \in \mathbb{Z}^{\leqslant 0}$ with $a-b \ne - (1+q)$, and $b \in \mathbb{Z}^{> 0}$}

Of course, $a \in \mathbb{Z}^{\leqslant 0}$ and $b \in \mathbb{Z}^{> 0}$ precludes $a-b \ne - (1+q)$.  Thus, this case does not occur.
\\

\noindent \underline{Case 4.A.  $a \in \mathbb{Z}^{\leqslant0}$ with $a-b = - (1+q)$, and $b \not\in \mathbb{Z}$}

Since $a \in \mathbb{Z}^{\leqslant0}$ and $b\not\in \mathbb{Z}$ precludes $a-b = - (1+q)$, this case does not occur.
\\

\noindent \underline{Case 4.B.  $a \in \mathbb{Z}^{\leqslant0}$ with $a-b = - (1+q)$, and $b \in \mathbb{Z}^{\leqslant0}$}

With $a = - m$ and $b = - n$, the constraint on $a-b$ implies $m > n$.  Thus, this case complements Case~3.B.

$b \in \mathbb{Z}^{\leqslant0} \Longrightarrow M(a,b,z)$ is not defined.  Moreover, the constraint $a-b = - (1+q) \Longrightarrow b = 1+a+q$, so that according to 13.2.8 of DLMF, $U(a,b,z) \propto \wt$.  Thus, out of our three standard solutions, we have only one remaining, either $U(a,b,z)$ or $\wt$. 

For $U(a,b,z)$, we use 13.2.7 or 13.2.8 of DLMF.  

We take the first solution to be $\widetilde{M}(-m,-n,z)$, although we could use $U(a,b,z)$.

Moreover, since $m \geqslant 1+n$, 13.2.31 of DLMF  allows us to take as our second solution
\begin{align}
&z^{n+1}\Bigg\{ \sum_{s\,=\,1}^{n+1} \frac{(n+1)!(s-1)!}{(n-s+1)!(m-n)_s}z^{-s}   \notag \\ &- \sum_{s\,=\,0}^{m-n-1} \frac{(-m+n+1)_s}{(n+2)_s}\frac{z^s}{s!}[\ln z + \psi(m-n-s) - \psi(1+s)- \psi(2+s+n)]  \notag \\ &+ (-1)^{m+n}(m-n-1)!\sum_{s\,=\,m\,-\,n}^{\infty} \frac{(-m+n+s)!}{(n+2)_s}\frac{z^s}{s!}\Bigg\} , \quad m \geqslant 1+ n.
\end{align}
Note the presence of the $\ln z$ terms.  Note that \textit{this is not} $U(-m,-n,z)$.

This is the first case where we have needed to go beyond the three standard solutions to obtain a second linearly independent solution.
\\

\noindent \underline{Case 4.C.  $a \in \mathbb{Z}^{\leqslant0}$ with $a-b = - (1+q)$, and $b \in \mathbb{Z}^{>0}$}

With $a = - m$ and $b = 1+n$, where $m \in \mathbb{Z}^{\geqslant 0}$ and $n \in \mathbb{Z}^{\geqslant 0}$, we have $a-b = -1-(m+n)$, and so $a-b = - (1+q)$ is ensured.

According to 13.2.7 of DLMF, $a \in \mathbb{Z}^{\leqslant0}$, $b\not\in \mathbb{Z}^{\leqslant0} \Longrightarrow U(a,b,z) \propto M(a,b,z)$.  Moreover, $b \in \mathbb{Z}^{>0} \Longrightarrow$ for $b = 1, M(a,b,z) = \wt$ and for $b \in \mathbb{Z}^{\geqslant 2}, \wt$ is not defined.  So we have only one distinct solution left from our three usual solutions, either $M(a,b,z)$ or $U(a,b,z)$.

We could get $U(a,b,z)$ from 13.2.7 or 13.2.10 of DLMF,

We take $M(-m,1+n,z)$ as our first solution, although we could use $U(-m,1+n,z)$.

Since $a = - m$ and $b = 1 = n$, we can use 13.2.28 of DLMF,
\begin{align}
&\sum_{s\,=\,1}^n \frac{n!(s-1)!}{(n-s)!(1+m)_s}z^{-s} \notag  \\ &- \sum_{s\,=\,0}^m \frac{(-m)_s}{(1+n)_s}\frac{z^s}{s!}[\ln z + \psi(1+m-s) - \psi(1+s)- \psi(1+s+n)] \notag \\ &+ (-1)^{1+m}m!\sum_{s\,=\,1+m}^{\infty} \frac{(s-1-m)!}{(n+1)_s}\frac{z^s}{s!},
\end{align}
as our second solution.  Note the presence of the $\ln z$ terms.  Note that \textit{this is not} $U(-m,1+n,z)$.

This is the second case where we have needed to go beyond the three standard solutions to obtain a second linearly independent solution.
\\

\noindent \underline{Case 5.A. $a \in \mathbb{Z}^{>0}$ with $a-b \ne -(1+q)$, $b \not\in \mathbb{Z}$}

Obviously, $a \in \mathbb{Z}^{>0}$, $b \not\in \mathbb{Z}$ ensures $a-b \ne -(1+q)$.

It should be clear that all three of our usual solutions are valid.

Since $b \not\in \mathbb{Z}$ , we can obtain $U(a,b,z)$ from equation \eqref{Eq:U1} or \eqref{Eq:U2}, or 13.2.42 of DLMF or 13.1.3 of~AS.

Consequently, with $a = 1+m$, we can take $M(1+m,b,z)$ and $U(1+m,b,z)$ as our two solutions.  We could, of course, use any two of $M(1+m,b,z)$, $\wta(1+m,b,z)$, and $U(1+m,b,z)$.
\\

\noindent \underline{Case 5.B. $a \in \mathbb{Z}^{>0}$ with $a-b \ne -(1+q)$, $b \in \mathbb{Z}^{\leqslant0}$}

With $a = 1+m$ and $b = - n$, where $m \in \mathbb{Z}^{\geqslant 0}$ and $n \in \mathbb{Z}^{\geqslant 0}$, we have $a-b = 1+m+n$, and so the constraint is ensured.

As usual, $b \in \mathbb{Z}^{\leqslant0} \Longrightarrow M(a,b,z)$ is not defined.

$U(a,b,z)$ follows from 13.2.11 and 13.2.9, or 13.2.30, of DLMF, or with care by 13.1.7 and 13.1.6 of AS.  Note carefully that both gamma functions in 13.2.9 will be finite, and so there will be $\ln z$ terms.

Accordingly, we take $\wta(1+m,-n,z)$ and $U(1+m,-n,z)$ as our two solutions.

\noindent \underline{Case 5.C. $a \in \mathbb{Z}^{>0}$ with $a-b \ne -(1+q), b \in \mathbb{Z}^{>0}$}

Obviously, $a-b \ne -(1+q) \Longleftrightarrow a \geqslant b$.  Thus, this case complements Case 6.C.

$b \in \mathbb{Z}^{>0} \Longrightarrow$ we can dispense with $\wt$.

$U(a,b,z)$ can be obtained from 13.2.9  or 13.2.27 of DLMF, or 13.1.6 of AS.  Note that there will be~$\ln z$ terms.

We can put $a = 1+m$ and $b=1+n$, where $m \in \mathbb{Z}^{\geqslant 0}$ and $n \in \mathbb{Z}^{\geqslant 0}$, with $m \geqslant n$, and take $M(1+m,1+n,z)$ and $U(1+m,1+n,z)$ as our solutions.
\\

\noindent \underline{Case 6.A. $a \in \mathbb{Z}^{>0}$ with $a-b = -(1+q)$, $b \not\in \mathbb{Z}$}

Obviously, $a \in \mathbb{Z}^{>0}$ and $b \not\in \mathbb{Z}$ ensures $a-b \ne-(1+q)$, and so this case does not occur.
\\

\noindent \underline{Case 6.B. $a \in \mathbb{Z}^{>0}$ with $a-b = -(1+q)$, $b \in \mathbb{Z}^{\leqslant0}$}

Clearly $a \in \mathbb{Z}^{>0}$, $b \in \mathbb{Z}^{\leqslant0}$ ensures that $a-b = -(1+q)$ cannot be satisfied.  Thus, this case also does not occur.
\\

\noindent \underline{Case 6.C. $a \in \mathbb{Z}^{>0}$ with $a-b = -(1+q)$, $b \in \mathbb{Z}^{>0}$}

We see that $a-b = -(1+q)$ requires that $a \leqslant b-1$, or with $a = 1+m$ and $b=1+n$, where $m \in \mathbb{Z}^{\geqslant 0}$ and $n \in \mathbb{Z}^{\geqslant 0}$, we must have $m < n$.  Thus, this case complements Case 5.C.

As we know all too well, $b \in \mathbb{Z}^{>0} \Longrightarrow \wt$ can be dispensed with.

$U(a,b,z)$ is given by 13.2.9 or 13.2.29 of DLMF, or 13.1.6 of AS.  Since $\Gamma(a-n) = \Gamma(m-n+1) \to \infty$, even though $b \in \mathbb{Z}$, there are no $\ln z$ terms in $U(a,b,z)$.

Thus, with $a = 1+m, b=1+n$, we can take $M(a,b,z)$ and $U(a,b,z)$ as our two solutions.
\\

As we have noted, all of these results are summarized in table~\ref{Table:Labyrinth}, which can be found in section \ref{S:Section3}.


 \ukrainianpart
 
 \title{ Довідник фізика з розв'язування рівняння Кумера та конфлюентних гіпергеометричних функцій }
 
 \author{У. Н. Метьюз мол.\refaddr{label1}, M. A. Езрік\refaddr{label1}, З. Тео\refaddr{label2}, Дж. K. Фрірікс \refaddr{label1} }
 
 \addresses{
	\addr{label1} Фізичний факультет, Джоржтаунський університет, північно-західні 37-а і O вул., Вашінгтон, округ Колумбія, США 20057-0995
	\addr{label2} Математичний факультет, Сяменьський університет Малайзії, Джалан Сунсурія, Бандар Сунсурія, Сепанг, 43900, Селанго, Малайзія
 }

 \makeukrtitle
 \begin{abstract}
	Конфлюентне гіпергеометричне рівняння, також відоме як рівняня Кумера, є одним з найважливіших диференціальних рівнянь фізики, хімії та  інженерних дисциплін. 
Його двома поліномними розв'язками є функція Кумера $M(a,b,z)$, яку часто називають  конфлюентною  гіпергеометричною функцією першого роду, а також $\widetilde{M}(a,b,z) \equiv z^{1-b}\;M(1+a-b,2-b,z)$, де $a$ і $b$ -- параметри, що входять у диференціальне рівняння.  
Зазвичай використовують також і третю функцію (функцію Трікомі), $U(a,b,z)$, яку деколи називають конфлюентною гіпергеометричною функцією другого роду, яка також є розв'язком конфлюентного гіпергеометричного рівняння. 
На відміну від загальноприйнятої практики, при пошуку двох лінійно незалежних розв'язків конфлюентного  гіпергеометричного рівняння слід розглядати як мінімум усі ці три функції. 
Існують ситуації, коли $a, b$ і $a - b$ є цілими числами, де одна з цих функцій є невизначеною, або дві з цих функцій не є лінійно незалежними, або один з лінійно незалежних розв'язків цього диференціального рівняння відрізняється від цих трьох функцій. 
Багато таких особливих випадків є в точності такими, які виникають при розв'язування фізичних задач. 
Це все призводить до великих непорозумінь щодо того, як саме слід підходити до розв'язування конфлюентних гіпергеометричних рівнянь, незважаючи на наявність авторитетних довідкових джерел, таких як цифрова бібліотека математичних функцій Національного інституту стандартів і технологій. У даній статті ми коректно описуємо усі ці випадки, а також те, якими є явні формули для двох лінійно незалежних розв'язків конфлюентного гіпергеометричного рівняння. Процедуру коректного розв'язування конфлюентного гіпергеометричного рівняння підсумовано у вигляді зручної таблиці. В якості прикладу ці розв'язки використано для дослідження зв'язаних станів воднеподібного атома, виправляючи стандартний підхід, описаний у підручниках. Ми також коротко розглядаємо обрізаний кулонівський потенціал.  Ми сподіваємося, що викладені методики будуть корисними для фізиків при розв'язуванні задач, в яких виникає конфлюентне гіпергеометричне диференціальне рівняння.

 	
 \end{abstract}

\lastpage 
\end{document}